\documentclass[aps,showpacs,nofootinbib]{revtex4}
\usepackage{amsmath}    
\usepackage{graphicx}   
\usepackage{color}      
\usepackage{amssymb}
\usepackage{graphics}
\usepackage{dcolumn}
\usepackage{bm}
\usepackage{epstopdf}

\begin{document}

    \newcommand{\blue}[1]{\textcolor{blue}{#1}}
    \newcommand{\new}{\blue}
    \newcommand{\green}[1]{\textcolor{green}{#1}}
    \newcommand{\modif}{\green}
    \newcommand{\red}[1]{\textcolor{red}{#1}}
    \newcommand{\attention}{\red}

    \title{Optical Theorem, Crossing Property and Derivative Dispersion Relations:
        Implications on the Asymptotic Behavior of $\sigma_{\footnotesize\mbox{tot}}(s)$ and $\rho(s)$}

    \author{S. D. Campos} \email{sergiodc@ufscar.br}
    \affiliation{Applied Mathematics Laboratory-CCTS/DFQM, \\Federal University of S\~ao Carlos, Sorocaba CEP 18052-780, Brazil}

    \author{V. A. Okorokov} \email{VAOkorokov@mephi.ru; Okorokov@bnl.gov}
    \affiliation{National Research Nuclear University MEPhI (Moscow
        Engineering Physics Institute), Kashirskoe highway 31, 115409
        Moscow, Russia}

    \date{\today}

    \begin{abstract}

        In this paper, one presents some results concerning the behavior
        of the total cross section and $\rho$-parameter at asymptotic
        energies in proton--proton ($pp$) and antiproton--proton
        ($\bar{p}p$) collisions. For this intent, we consider three of the
        main theoretical results in high energy physics: the crossing
        property, the derivative dispersion relation, and the optical
        theorem. The use of such machinery allows the analytic formulas
        for wide set of the measured global scattering parameters and some
        important relations between them. The suggested parameterizations
        approximate simultaneously the energy dependence for total cross
        section and $\rho$-parameter for $pp$ and $\bar{p}p$ with
        statistically acceptable quality in multi-TeV region. Also the
        qualitative description is obtained for important interrelations,
        namely difference, sum and ratio of the antiparticle--particle and
        particle--particle total cross sections. Despite the reduced
        number of experimental data for the total cross section and
        $\rho$-parameter in TeV-scale, which turns any prediction for the
        beginning of the asymptotic domain a hard task, the fitting
        procedures indicates that asymptotia lies in the energy range
        25.5--130 TeV. Moreover, in the asymptotic regime, one obtains
        $\alpha_{\mathbb{P}}=1$. Detailed quantitative study of energy
        behavior of measured scattering parameters and their combinations
        in ultra--high energy domain indicates that the scenario with the
        generalized formulation of the Pomeranchuk theorem is more
        favorable with respect to the original formulation of this
        theorem.

    \end{abstract}


    \maketitle

    \section{Introduction}\label{sec:intro}

    The absence of a rigorous formalism able to predict elastic and
    diffractive processes, the so-called soft scattering states, turns
    soft interactions into a great challenge for quantum
    chromodynamics (QCD). As a complementary difficulty, the use of
    asymptotic theorems in high energy physics are also a hard matter
    since there is no indication when they must start to be applied,
    i.e. where the asymptotia should begins. Furthermore, they also
    emerge in different contexts in the $S$-matrix, a pre-QCD
    formalism.

    There are a few of the rigorous theorems which are crucially
    important for high energy physics, and, especially, for the
    asymptotic energy domain. In the present work, we analyze two of
    the most outstanding among them, namely Froissart--Martin bound
    and Pomeranchuk theorem, both concerning particle--particle and
    antiparticle--particle total cross sections
    ($\sigma_{\footnotesize\mbox{tot}}$), a forward quantity (zero
    transferred momentum).

    Probably, the so-called Froissart--Martin bound
    \cite{froissart_phys_rev_123_1053_1961,martin_nuovo_cim_42_930_1966,A.Martin.Nuovo.Cim.A44.1219.1966,lukaszuk_nuovo_cimen_a52_122_1967}
    and its recent modification \cite{Martin-PR-D80-065013-2009}, is
    the most relevant asymptotic bound in forward high energy
    scattering, since it furnishes some kind of physical limit
    to the total cross section rise as the collision energy grows.
    This theoretical result is the effect of the analyticity
        of elastic scattering amplitude and rigorous consequence of the
        most general properties of quantum field theory (QFT), namely,
        causality, unitarity, and the polynomial boundness
        \cite{martin_nuovo_cim_42_930_1966,A.Martin.Nuovo.Cim.A44.1219.1966}.
        This asymptotic bound is pre-QCD and works, in fact, as a general
    parameterization for several models concerning the rise of
    $\sigma_{\footnotesize\mbox{tot}}$. In the QCD framework, there is
    no formal derivation of this bound from the first principles
    unless the functional integral approach
    \cite{O.Nachtmann.Ann.Phys.209.436.1991}. However, the absence of
    a non-perturbative QCD corroboration does not contradict the
    robustness of such inequality.

    Another interesting forward asymptotic result is given by the
    Pomeranchuk theorem \cite{pomeranchuk_jetp_7_499_1958}, and it
    concerns the difference between $\sigma_{\footnotesize\mbox{tot}}$
    for particle--particle and antiparticle--particle. As
        well-known, this theorem is the consequence of the crossing
        property and an effect of the analyticity of the scattering amplitude.
        The Pomeranchuk theorem predicts that, for sufficiently high
    energies, the difference between these total cross sections should
    vanish. The general belief about this result imposes the need for
    a particle exchange as responsible for the vanishing difference as
    the collision energy grows. This particle is called the leading
    Regge pole \cite{S.D.Campos.Chin.Phys.C.2020}, or Pomeron for
    short, and it does not differentiate particle from antiparticle
    since it possesses the vacuum quantum numbers. The usual picture
    where the Pomeron is viewed as a pair of gluons is due to Low and
    Nussinov
    \cite{F.E.Low.Phys.Rev.D12.163.1975,S.Nussinov.Phys.Rev.Lett.34.1286.1975}.
    Despite the huge amount of experimental data, there is no evidence
    for the Pomeron up to the present-day energies.

    In the phenomenological context, the above two theorems remain as
    fundamental to impose constraints on the rise of
    $\sigma_{\footnotesize\mbox{tot}}$ as the collision energy grows.
    However, they do not stand alone when we talk about forward
    quantities. The optical theorem is a remarkable result connecting
    the total cross section and the imaginary part of the forward
    scattering amplitude \cite{eden_book_2}. Moreover, the derivative
    dispersion relations (DDR) can be used to connect the
    imaginary part of the forward scattering amplitude with the real
    part \cite{S.D.Campos.EPJC.47.171.2006}. Thus, at least in the
    forward collision context, a whole scattering amplitude can be
    constructed based on few theoretical results.

    It should be stressed that if we assume that crossing property is
    valid, despite its lack of theoretical and experimental evidence,
    then we can obtain the forward scattering amplitude for
    particle--particle events from the forward scattering amplitude
    for antiparticle--particle scattering (and vice--versa).

    In the present work, one uses the crossing property, the
    DDR, and the optical theorem to obtain
    some theoretical results on the rise of
    $\sigma_{\footnotesize\mbox{tot}}$. In particular, based on these
    results, one presents a simple fitting model for the
    proton--proton ($pp$) and antiproton--proton ($\bar{p}p$) total
    cross sections. Our results indicate that, asymptotically, a
    Pomeron intercept $\alpha_{\mathbb{P}}=1$. Moreover, one also
    estimates the energy range where the asymptotic regime should
    begins.

    The paper is organized as follows. In section \ref{sec:model},
    some considerations are discussed about the total cross section.
    Section \ref{sec:expDB} contains a detailed description of
    experimental databases considered in the paper and corresponding
    procedures for approximation. Section \ref{sec:exp-1} presents the
    parameters obtained in the fitting procedure. Discussion and
    projections for some global scattering parameters are in the
    Section \ref{sec:5}. Section \ref{sec:6} contains conclusions and
    final remarks.

    \section{Asymptotic Behavior of Total Cross Section and $\rho$-Parameter}\label{sec:model}

    As is well-known, the crossing property has never been
    proven (see, for instance, \cite{Collins-book-1968}), but in general, based on it, one postulates that
    scattering amplitudes can be analytically continued between the
    different channels of the collision process. Then, one assumes
    that it is possible to write the scattering amplitude in terms of
    auxiliary even ($+$) and odd ($-$) amplitudes as
    \begin{eqnarray}\label{eq:2.1}
        \displaystyle
        2f_\pm(s,t)=F_{pp}(s,t) \pm F_{\bar{p}p}(s,t),
    \end{eqnarray}

    \noindent where
    $f_\pm(s,t)=\mathrm{Re}f_\pm(s,t)+i\mathrm{Im}f_\pm(s,t)$ are the
    crossing amplitudes, and $F_{pp}(s,t)$ and $F_{\bar{p}p}(s,t)$ are
    the complex-valued scattering amplitudes for $pp$ and $\bar{p}p$
    processes. As usual, $s$ stands for the square of collision energy
    and $-t$ for the square of transferred momentum, both in the
    center--of--mass system.

    Based on the above definition, the DDR can be written, up to
    the first-order approximation, for the odd and even amplitudes
    (\ref{eq:2.1}) in the forward direction ($t=0$) as
    \cite{NPA-744-249-2004,S.D.Campos.EPJC.47.171.2006}
    \begin{subequations}
        \begin{equation}
            \displaystyle
            \frac{\mathrm{Re}f_+(s)}{s}=\frac{k}{s}+\biggl[\frac{\pi}{2}\frac{d}{d\ln s}\biggr]\frac{\mathrm{Im}f_+(s)}{s}, \label{eq:2.2a}
        \end{equation}
        \begin{equation}
            \displaystyle
            \frac{\mathrm{Re}f_-(s)}{s}=\biggl[\frac{\pi}{2}\biggl(1+\frac{d}{d\ln s}\biggr)\biggr]\frac{\mathrm{Im}f_-(s)}{s}, \label{eq:2.2b}
        \end{equation}\label{eq:2.2}
    \end{subequations}

    \noindent where $f_{\pm}(s,0) \equiv f_{\pm}(s)$, and $k$ is the
    subtraction constant. Without loss of generality, one adopts $k=0$
    since the influence of such parameter is restricted to the low
    energy domain \cite{S.D.Campos.Chin.Phys.C.2020}. Notice the
    addition of high-order derivative terms in equations
    (\ref{eq:2.2}) may turn this representation more sensitive to
    describe high-energy experimental data. Taking into account the
    representation (\ref{eq:2.1}), and DDR (\ref{eq:2.2}), the
    simple scheme below summarizes the use of such odd and even
    amplitudes
    \begin{eqnarray}
        \displaystyle
        \nonumber   \mathrm{Im}F_{xp}(s)\longrightarrow \mathrm{Im}f_{\pm}(s)\longrightarrow \mathrm{Re}f_{\pm}(s)\longrightarrow \mathrm{Re}F_{xp}(s),
    \end{eqnarray}

    \noindent where $x=p, \bar{p}$. This scheme reveals the importance
    of phenomenological information (the input) about the imaginary
    part of the forward scattering amplitude. Furthermore, through the
    optical theorem, the forward imaginary part can be used to define
    the total cross section behavior. As is well-known, the
    optical theorem lives in the core of high energy physics, and it
    was proved to be valid for all energies and scattering
    angles \cite{eden_book_2}, being written as
    \begin{eqnarray}\label{eq:asymp_6}
        \displaystyle
        \sigma_{\footnotesize\mbox{tot}}^{xp}(s)=\mathrm{Im}F_{xp}(s) / s.
    \end{eqnarray}

    The optical theorem (\ref{eq:asymp_6}), as well as the DDR
    (\ref{eq:2.2}), can be used to obtain the two following results
    concerning the behavior of $\sigma_{\footnotesize\mbox{tot}}$ as
    $s$ grows
    \begin{subequations}
        \begin{equation}
            \displaystyle
            \frac{d\Delta_{\footnotesize\mbox{tot}}(s)}{d\ln s}=
            \frac{2}{s}\left\{\frac{1}{\pi}\bigl[\mathrm{Re}F_{\bar{p}p}(s)-\mathrm{Re}F_{pp}(s)\bigr]+\mathrm{Im}f_-(s)\right\}, \label{eq:2.4a}
        \end{equation}
        \begin{equation}
            \displaystyle
            \frac{d\Sigma_{\footnotesize\mbox{tot}}(s)}{d\ln s}=\frac{2}{\pi s}\bigl[\mathrm{Re}F_{\bar{p}p}(s)+\mathrm{Re}F_{pp}(s)\bigr]
            \label{eq:2.4b}
        \end{equation}\label{eq:2.4}
    \end{subequations}

    \noindent  where the difference and sum of $pp$ and $\bar{p}p$
    total cross sections are written, respectively, as
    \begin{subequations}
        \begin{equation}
            \displaystyle \Delta_{\footnotesize\mbox{tot}}(s) \equiv
            \sigma_{\footnotesize\mbox{tot}}^{\bar{p}p}(s) -
            \sigma_{\footnotesize\mbox{tot}}^{pp}(s), \label{eq:2.5a}
        \end{equation}
        \begin{equation}
            \displaystyle \Sigma_{\footnotesize\mbox{tot}}(s) \equiv
            \sigma_{\footnotesize\mbox{tot}}^{\bar{p}p}(s) +
            \sigma_{\footnotesize\mbox{tot}}^{pp}(s). \label{eq:2.5b}
        \end{equation}\label{eq:2.5}
    \end{subequations}

\noindent One can note the relation (\ref{eq:2.4a}) implies the independence of $\Delta_{\footnotesize\mbox{tot}}(s)$ and $\mathrm{Im}f_{-}(s)$ on each other for further calculations which in turn is based on the above definitions of the parameters $\Delta_{\footnotesize\mbox{tot}}(s)$ and $\Sigma_{\footnotesize\mbox{tot}}(s)$ via only measuring quantities ($\sigma_{\footnotesize\mbox{tot}}^{xp}$) without possible consequent transition to the imaginary parts of amplitudes. Of course, the experimentally measured $pp$ and
    $\bar{p}p$ total cross sections can be written using the above
    results as follow:
    \begin{subequations}
        \begin{equation}
            \displaystyle
            2\sigma_{\footnotesize\mbox{tot}}^{\bar{p}p}(s)=\Sigma_{\footnotesize\mbox{tot}}(s)
            + \Delta_{\footnotesize\mbox{tot}}(s), \label{eq:2.6a}
        \end{equation}
        \begin{equation}
            \displaystyle
            2\sigma_{\footnotesize\mbox{tot}}^{pp}(s)=\Sigma_{\footnotesize\mbox{tot}}(s)
            - \Delta_{\footnotesize\mbox{tot}}(s). \label{eq:2.6b}
        \end{equation}\label{eq:2.6}
    \end{subequations}

    Equations (\ref{eq:2.4}) can be analyzed taking into account some
    expected particularities for $\mathrm{Im}F_{xp}(s)$ and
    $\mathrm{Re}F_{xp}(s)$. First of all, at very high energies one
    expects, from the phenomenological point of view, that
    \begin{eqnarray}\label{eq:real_0}
        \displaystyle
        \mathrm{Re}F_{xp}(s) \ll \mathrm{Im}F_{xp}(s),
    \end{eqnarray}

    \noindent which means an $almost$ complete absorptive scattering.
    On the other hand, the Froissart--Martin bound states that for
    $pp$ and $\bar{p}p$ forward collision, the total cross section
    obeys the inequality
    \begin{eqnarray}\label{eq:asymp_2}
        \left.\sigma_{\footnotesize\mbox{tot}}(s)\right|_{s \to \infty} \leq C\ln^2\varepsilon,
    \end{eqnarray}

    \noindent where $C$ is a constant, $\varepsilon \equiv s/s_0$ and
    $s_0$ is some fixed scale which is, in general, unspecified. The
    scale can be chosen to hadronic particles as the reasonable one
    $s_0=1$ GeV$^{2}$ \cite{Donnachie-book-2002}. At this choice of
    $s_{0}$, the axiomatic quantum field theory (AQFT) resulted in
    $C=\pi/m_{\pi}^2\approx 62.8$ mb
    \cite{A.Martin.Nuovo.Cim.A44.1219.1966,lukaszuk_nuovo_cimen_a52_122_1967}
    while the twice smaller value $\pi/2m_{\pi}^2\approx 31.4$ mb was
    recent derived \cite{Martin-PR-D80-065013-2009} with
        $m_{\pi}$ being the charged pion mass
    \cite{PTEP-2020-083C01-2020}. It is well-known that bound
    (\ref{eq:asymp_2}) cannot be improved using only analyticity in
    the momentum transfer, unitarity, and boundedness by a polynomial
    in $s$, even if oscillations were allowed
    \cite{T.Kinoshita.J.J.Loeffel.A.Martin.Phys.Rev.Lett.10.460.1963}.

    The theoretical results (\ref{eq:asymp_6}) and (\ref{eq:asymp_2})
    allow the construction of a wide room to accommodate several
    functional forms of $\mathrm{Re}F_{xp}(s)$ which satisfies the
    phenomenological condition (\ref{eq:real_0}). The simplest
    suggestion among them is to choose a sufficiently small constant
    (which can always be done), in order to satisfy relation
    (\ref{eq:real_0}) at asymptotically high energies. Then, one can
    suppose from some energy $\sqrt{s_a}$, and in consideration of the
    onset of the asymptotic region in the elastic scattering case
    that: (i) $\mathrm{Re}F_{xp}(s)=0$, or (ii)
    $\mathrm{Re}F_{xp}(s)/s=c_x\neq 0$ is a small real number
    depending on $x$, or (iii) $\mathrm{Re}F_{xp}(s)/s=c\neq 0$, where
    $c$ is a small real number.

    Taking into account the assumption (i) in (\ref{eq:2.2b}), one
    obtains a simple expression for the odd auxiliary function
    \begin{eqnarray}\label{eq:asympt_1}
        \displaystyle
        \mathrm{Im}f_-(s) / s = a_{1}/\varepsilon,
    \end{eqnarray}

    \noindent where $a_1$ is an integration constant and, for the sake
    of simplicity, hereafter one adopts the lower limit of integration
    as $s_0$. Using result (\ref{eq:asympt_1}) and assumption (i),
    then from (\ref{eq:2.4a}) one has
    \begin{eqnarray}\label{eq:res_1}
        \displaystyle
        \Delta_{\footnotesize\mbox{tot}}(s)= -2a_{1}/\varepsilon.
    \end{eqnarray}

    Notice that the sign of $a_1$ determines which of both total cross
    sections rise faster as $s$ grows. For example, if $a_1<0$, one
    has
    $\sigma_{\footnotesize\mbox{tot}}^{\bar{p}p}(s)\gtrsim\sigma_{\footnotesize\mbox{tot}}^{pp}(s)$,
    for asymptotic energies.

    At the end of the 1950s, the general belief assure that total
    cross section decreases with the increasing collision energy, as
    shown by the experimental data. As is well-known, this
    conviction was only modified with the Intersecting Storage Rings
    (ISR) measurement of the $pp$ total cross section done in 1973
    showing the increase of $\sigma_{\footnotesize\mbox{tot}}^{pp}(s)$
    with energy.

    Possibly, the first theoretical result using the asymptotic
    condition as a way to obtain some useful ground in high energy
    physics is the Pomeranchuk theorem
    \cite{pomeranchuk_jetp_7_499_1958}. The original version assumes
    that if the forward elastic scattering amplitude grows slower than
    $s$, then the difference between $\bar{p}p$ and $pp$ total cross
    sections tend to zero, as $s\rightarrow\infty$
    \begin{eqnarray}\label{eq:asymp_1}
        \Delta_{\footnotesize\mbox{tot}}(s) \rightarrow 0, ~\mathrm{if}~ |F(s)|<s.
    \end{eqnarray}

    The Pomeranchuk proof uses dispersion relations and some
    additional intuitive assumptions, removed in other versions of the
    Pomeranchuk result, usually written as
    \cite{eden_phys_rev_lett_16_39_1966,kinoshita_book_1966,grunberg_phys_rev_lett_31_63_1973}
    \begin{eqnarray}\label{eq:asymp_1b}
        R_{\footnotesize\mbox{tot}}^{\bar{p}/p}(s) \equiv \sigma_{\footnotesize\mbox{tot}}^{\bar{p}p}(s) / \sigma_{\footnotesize\mbox{tot}}^{pp}(s) \rightarrow 1, ~\mathrm{if}~ s \rightarrow \infty,
    \end{eqnarray}

    \noindent which, one stresses, is not the same result expressed by (\ref{eq:asymp_1}).

    In the light of the present assumption (i), when
    $s\rightarrow\infty$ the result (\ref{eq:res_1}) vanishes
    asymptotically, corroborating the Pomeranchuk theorem. In
    contrast, if we use the assumption (ii), it implies in the
    following result
    \begin{eqnarray}\label{eq:real_diff_1}
        \displaystyle
        \bigl[\mathrm{Re}F_{\bar{p}p}(s)-\mathrm{Re}F_{pp}(s)\bigr] / s=
        c_{\bar{p}}-c_p.
    \end{eqnarray}

    \noindent In this case, one has from (\ref{eq:2.2b})
    \begin{eqnarray}\label{eq:asympt_2}
        \displaystyle
        \mathrm{Im}f_{-}(s) / s = (c_{\bar{p}}-c_p)/\pi+a_{2}/\varepsilon.
    \end{eqnarray}

    \noindent where $a_2\neq 0$ is an integration constant. Using
    above result, then the difference between the total cross sections
    is written as
    \begin{eqnarray}\label{eq:res_2}
        \displaystyle
        \Delta_{\footnotesize\mbox{tot}}(s)= \bigl[4(c_{\bar{p}}-c_p)/\pi \bigr]\ln \varepsilon - 2a_{2}/\varepsilon,
    \end{eqnarray}

    \noindent which does not corroborate the Pomeranchuk theorem
    unless $c_{\bar{p}}=c_p$. The last assumption (iii) implies
    $c_{\bar{p}}=c_p=c$, and, in this case, the difference is null
    \begin{eqnarray}
        \displaystyle
        \mathrm{Re}F_{\bar{p}p}(s)-\mathrm{Re}F_{pp}(s) = 0.
    \end{eqnarray}

    The above result implies an asymptotic behavior for the
    total cross section similar to (\ref{eq:res_1}). From this simple
    analysis, it is possible to conclude that $\mathrm{Re}F_{xp}(s)=0$
    or $\mathrm{Re}F_{xp}(s)/s=c$ at high energies since both results
    corroborate with the Pomeranchuk theorem.

    Equation (\ref{eq:2.4b}) can also provide physical information on
    the behavior of $\Sigma_{\footnotesize\mbox{tot}}(s)$. Taking into
    account assumption (i), then from some $s_a$, one has
    \begin{eqnarray}\label{eq:res_3a}
        \Sigma_{\footnotesize\mbox{tot}}(s)= a_{0},
    \end{eqnarray}

    \noindent where $a_{0}$ is a real constant. It is important to
    stress this result does not violate any theorem in high energy
    physics, and it seems to indicate the existence of some taming
    mechanism (for example, the mini-jet
    \cite{D.A.Fagundes.A.Grau.G.Pancheri.O.Shekhovtsova.Y.N.Srivastava.Phys.Rev.D96.054010.2017})
    to the rise of $\sigma_{\footnotesize\mbox{tot}}(s)$ as $s$ grows.

    In contrast, for a non-null real part given by assumption (ii),
    one obtains
    \begin{eqnarray}\label{eq:res_4}
        \displaystyle
        \Sigma_{\footnotesize\mbox{tot}}(s)= \bigl[2(c_{\bar{p}}+c_p)/\pi \bigr]\ln \varepsilon,
    \end{eqnarray}

    \noindent which notably reveals that in the asymptotic limit the
    sum of the cross sections follow the logarithm of the collision
    energy, i.e. with a Pomeron intercept $\alpha_{\mathbb{P}}=1$.
    Considering assumption (iii), one can derive a similar result
    \begin{eqnarray}\label{eq:res_5}
        \displaystyle
        \Sigma_{\footnotesize\mbox{tot}}(s)= (4c/\pi)\ln \varepsilon.
    \end{eqnarray}

    It is unnecessary to say that results (\ref{eq:res_4}) and
    (\ref{eq:res_5}) does not violate the Froissart--Martin theorem
    (\ref{eq:asymp_2}) even in its modified version
    \cite{Martin-PR-D80-065013-2009}.

    For clarity, all the above results are summarized in Table
    \ref{tab:1}. From this table, one observes that for (i), although
    $\Delta_{\footnotesize\mbox{tot}}(s)$ obeys the Pomeranchuk
    theorem, the sum $\Sigma_{\footnotesize\mbox{tot}}(s)$ seems to
    not correspond to the behavior shown by the experimental data, at
    least in the present-day energies. For the results expressed by
    (ii), on the other hand, one has that
    $\Delta_{\footnotesize\mbox{tot}}(s)$ does not obey the
    Pomeranchuk theorem, unless $c_{\bar{p}}=c_p$. The last results
    given by (iii) seems to be reasonable under the lights of the
    experimental data, representing a Pomeron intercept
    $\alpha_{\mathbb{P}}=1$ \cite{S.D.Campos.Chin.Phys.C.2020}.

    \begin{table}[h!]
        \caption{Summary of the theoretical results
            obtained assuming the asymptotic condition $s\rightarrow \infty$.}\label{tab:1}
        \begin{center}
            \begin{tabular}{ccccc}
                \hline
                \multicolumn{1}{c}{Assumption} & \multicolumn{1}{c}{$\mathrm{Re}F_{xp}(s)/s$} &
                \multicolumn{1}{c}{$\mathrm{Im}f_{-}(s)/s$} & \multicolumn{1}{c}{$\Delta_{\footnotesize\mbox{tot}}(s)$} &
                \multicolumn{1}{c}{$\Sigma_{\footnotesize\mbox{tot}}(s)$}  \rule{0pt}{10pt}\\
                \hline
                (i) & 0 & ~$a_1/\varepsilon$~ & ~$-2a_1/\varepsilon$~ & ~$a_{0}$ \rule{0pt}{10pt}\\
                \hline
                (ii) & $c_x$ & ~$(c_{\bar{p}}-c_p)/\pi+a_2/\varepsilon$~ &
                ~$[4(c_{\bar{p}}-c_p)/\pi]\ln \varepsilon-2a_2/\varepsilon$~ &
                ~$[2(c_{\bar{p}}+c_p)/\pi]\ln \varepsilon$ \rule{0pt}{10pt}\\
                \hline
                (iii)& $c$ & ~$a_1/\varepsilon$~ & ~$-2a_1/\varepsilon$~ & ~$(4c/\pi)\ln \varepsilon$~  \rule{0pt}{10pt}\\
                \hline
            \end{tabular}
        \end{center}
    \end{table}

    Therefore, one may conclude that $pp$ and $\bar{p}p$ real part of
    the forward elastic amplitude are null above some energy
    $\sqrt{s_a}$ or that they are equals. Both conclusions preserve
    the Pomeranchuk theorem as well as the Froissart--Martin bound.
    However, they lead to different asymptotic behaviors for
    $\Sigma_{\footnotesize\mbox{tot}}(s)$, which represents a puzzle
    that apparently cannot be solved at present-day energy. It is
    important to stress that considering the results shown in Table
    \ref{tab:1} as well as (\ref{eq:2.6}), one can obtain analytic
    expressions for experimentally measured quantities
    $\sigma_{\footnotesize\mbox{tot}}^{\bar{p}p}(s)$ and
    $\sigma_{\footnotesize\mbox{tot}}^{pp}(s)$ at asymptotically high
    energies.

    Now, we can apply the above procedure to study the behavior of the
    $\rho$-parameter as the collision energy grows. This parameter is
    defined as
    \begin{eqnarray}\label{eq:rho_1}
        \displaystyle \rho^{xp}(s)=\mathrm{Re}F_{xp}(s) /\,
        \mathrm{Im}F_{xp}(s),
    \end{eqnarray}

    \noindent which measures the rise of the absorptive part of the
    forward scattering amplitude (relative to the real part) as $s$
    grows.

    If the condition (i) is used, then, from some $s_{a}$, the only
    possible result is $\rho^{xp}(s_a < s)=0$, which does not
    contradict any theorem but has no predictive capability for non
    asymptotic energies. In contrast, adopting, for instance,
    assumption (iii), one obtains
    \begin{eqnarray}\label{eq:rho_2}
        \displaystyle
        \rho^{xp}(s)=c/\bigl[\pm a_1/\varepsilon+(2c/\pi)\ln \varepsilon \bigr],
    \end{eqnarray}

    \noindent where the sign $+/-$ is for $pp/\bar{p}p$. For some
    sufficient high energy, the term $a_1/\varepsilon$ can be
    disregarded, resulting the prediction that
    $\rho^{pp}(s)=\rho^{\bar{p}p}(s)$ and $\rho$-parameter approaches
    to its asymptotic value ($\rho_{a}$)
    \begin{eqnarray}\label{eq:rho_3}
        \displaystyle \left.\rho^{xp}(s)\right|_{s \to \infty} \to
        \rho^{xp}_{a}(s) \equiv (\pi /2)\ln^{-1} \varepsilon.
    \end{eqnarray}

    It is important to emphasize that (\ref{eq:rho_3}) is independent
    of any external parameter, being the collision energy the only
    variable of interest. It seems to be reasonable since we expect
    the same parameters that drive $\mathrm{Im}F_{xp}(s)/s$ should
    also control $\mathrm{Re}F_{xp}(s)/s$, at least for the high
    energy regime. Furthermore, the asymptotic result (\ref{eq:rho_3})
    agrees well with the ''standard" picture of high energy elastic
    diffraction in which amplitude $f_{-}(s)$ becomes negligible
    compared with the crossing-even one $f_{+}(s)$ as $s$ increases.
    The property of analyticity implies $\rho\propto
    \ln^{-1}\varepsilon$ \cite{Leader-book-V2-1996}.

    Results obtained for measured forward scattering parameters for
    $\bar{p}p$ and $pp$ within assumptions under consideration are
    summarized in Table \ref{tab:2}. It is interesting to point out
    that if we assume $\rho(s)$ as given by assumptions (i), (ii), or
    (iii), then the use of DDR conduces to result similar to
    those obtained in
    \cite{T.Kinoshita.J.J.Loeffel.A.Martin.Phys.Rev.Lett.10.460.1963}.
    In particular, assumptions (ii) and (iii) for $\rho(s)$ result in
    $\sigma_{\footnotesize\mbox{tot}}$ as given by Froissart--Martin
    bound, which means that it cannot be improved by the methods
    employed here.

    \begin{table}[h!]
        \caption{Summary of the results for measured forward
            parameters assuming the asymptotic condition $s\rightarrow \infty$.}\label{tab:2}
        \begin{center}
            \begin{tabular}{ccccc}
                \hline \multicolumn{1}{c}{Assumption} &
                \multicolumn{1}{c}{$\sigma_{\footnotesize\mbox{tot}}^{\bar{p}p}(s)$}
                & \multicolumn{1}{c}{$\sigma_{\footnotesize\mbox{tot}}^{pp}(s)$} &
                \multicolumn{1}{c}{$\rho^{\bar{p}p}(s)$} &
                \multicolumn{1}{c}{$\rho^{pp}(s)$}  \rule{0pt}{10pt}\\
                \hline
                (i) & ~$a_{0}/2-a_{1}/\varepsilon$~ & ~$a_{0}/2+a_{1}/\varepsilon$~ & ~0~ & ~0 \rule{0pt}{10pt}\\
                \hline (ii) & ~$\displaystyle \frac{3c_{\bar{p}}-c_{p}}{\pi}\ln
                \varepsilon-\frac{a_2}{\varepsilon}$~ & ~$\displaystyle \frac{3c_{p}-c_{\bar{p}}}{\pi}\ln
                \varepsilon+\frac{a_2}{\varepsilon}$~ & ~$\displaystyle
                \frac{c_{\bar{p}}}{[(3c_{\bar{p}}-c_p)/\pi]\ln
                    \varepsilon-a_2/\varepsilon}$~ & ~$\displaystyle
                \frac{c_{p}}{[(3c_{p}-c_{\bar{p}})/\pi]\ln
                    \varepsilon+a_2/\varepsilon}$ \rule{0pt}{15pt}\\
                \hline (iii)& ~$\displaystyle \frac{2c}{\pi}\ln \varepsilon - \frac{a_1}{\varepsilon}$~ &
                ~$\displaystyle \frac{2c}{\pi}\ln \varepsilon + \frac{a_1}{\varepsilon}$~ & ~$\displaystyle
                \frac{c}{(2c/\pi)\ln \varepsilon - a_1/\varepsilon}$~ &
                ~$\displaystyle \frac{c}{(2c/\pi)\ln \varepsilon + a_1/\varepsilon}$~  \rule{0pt}{15pt}\\
                \hline
            \end{tabular}
        \end{center}
    \end{table}

    In the framework of the present approach based on the crossing
    property, DDR (\ref{eq:2.2}), and optical theorem
    (\ref{eq:asymp_6}) both $pp$ and $\bar{p}p$ elastic collisions are
    characterized by similar energy dependencies for total cross
    section and $\rho$-parameter at asymptotically high energies,
    namely, $\forall\,x=p, \bar{p}:
    \sigma_{\footnotesize\mbox{tot}}^{xp}(s) \propto \ln \varepsilon$,
    $\rho^{xp} \propto \ln^{-1}\varepsilon$ within more realistic
    assumptions (ii) and (iii), which are considered in this paragraph
    below.

    It seems the dependence
    $\sigma_{\footnotesize\mbox{tot}}^{xp}(s)$ for $x=p, \bar{p}$ from
    Table \ref{tab:2} in the collision energy domain under consideration is
    functionally close to the increase of
    $\sigma_{\footnotesize\mbox{tot}}^{pp}(s)$ deduced within
    Regge--eikonal approach \cite{IJMPA-33-1850077-2018}. But one can
    note $\forall\,x=p, \bar{p}:
    \sigma_{\footnotesize\mbox{tot}}^{xp}(s)$ shows weaker increase
    with $s$ than that within AQFT and semiclassical color glass
    condensate (CGC) approach which imply
    $\sigma_{\footnotesize\mbox{tot}}(s)$ is functionally close to the
    Froissart--Martin limit (\ref{eq:asymp_2}) in the region, at
    least, of $\mathcal{O}$(100 TeV) energies \cite{PAN-81-508-2018}.
    The asymptotic behavior $\forall\,x=p, \bar{p}:
    \left.\rho^{xp}(s)\right|_{s \to \infty} \propto
    \ln^{-1}\varepsilon$ is qualitatively similar in functional sense
    to the corresponding AQFT $\rho(s)$ dependencies for $pp$ and
    $\bar{p}p$ taking into account the values of fit parameters
    \cite{IJMPA-A25-5333-2010,IJMPA-32-1750175-2017}.

    \section{Experimental Database and Fitting Procedure} \label{sec:expDB}

    The set of the global scattering parameters $\mathcal{G}_{1}
    \equiv \{\mathcal{G}_{1}^{i}\}_{i=1}^{4}=
    \{\sigma_{\scriptsize{\mbox{tot}}}^{pp},
    \sigma_{\scriptsize{\mbox{tot}}}^{\bar{p}p},\rho^{pp},\rho^{\bar{p}p}\}$
    contains only observables which are independent on each other as
    well as directly measured in experiments. On the other hand, the
    set $\mathcal{G}_{2} \equiv \{\mathcal{G}_{2}^{i}\}_{i=1}^{3}=
    \{\Delta_{\scriptsize{\mbox{tot}}},
    \Sigma_{\scriptsize{\mbox{tot}}},R_{\scriptsize{\mbox{tot}}}^{\bar{p}/p}\}$
    is formed by the parameters, strictly speaking, depending on
    experimentally measurable quantities and, moreover, terms of
    $\mathcal{G}_{2}$ are independent on each other as well as the
    terms of $\mathcal{G}_{1}$. In the present paper, the sets of
    scattering parameters and their combinations $\mathcal{G}_{j}$,
    $j=1,2$ are studied. Also the joined ensemble
    $\mathcal{G}=\mathcal{G}_{1} \bigcup\,\mathcal{G}_{2}$ is
    considered for completeness of information\footnote{In the paper
        total errors are used for experimental points unless otherwise
        specified. The total error is calculated as systematic error added
        in quadrature to statistical one.}.

    The experimental database for $\mathcal{G}_{1}$ contained the
    ensembles for $\sigma_{\scriptsize\mbox{tot}}^{xp}$, $\rho^{\,xp}$
    from \cite{PTEP-2020-083C01-2020} is denoted as DB20 while the
    database taken into account the above samples and results from
    STAR for $\sigma_{\scriptsize\mbox{tot}}^{pp}$ at $\sqrt{s}=0.20$
    TeV \cite{PLB-808-135663-2020} and from TOTEM for $\rho^{pp}$ at
    $\sqrt{s}=13$ TeV \cite{EPJC-79-785-2019} is referred as DB20+.
    The last paper leads to some uncertainty which result in two
    points of view for database creation. Two values
    $\left.\rho_{1}^{pp}\right|_{\sqrt{s}=13\,\scriptsize{\mbox{TeV}}}=0.09
    \pm 0.01$ and
    $\left.\rho_{2}^{pp}\right|_{\sqrt{s}=13\,\scriptsize{\mbox{TeV}}}=0.10
    \pm 0.01$ have been obtained in \cite{EPJC-79-785-2019} for one
    quantity and collision energy without any preference for one value
    of $\rho^{pp}$ on the another. On the other hand, the one result
    should be included in the corresponding data sample\footnote{It
        should be stressed that such request is fully within the rules
        were applied for creation of experimental databases in various
        analyzes, for instance, for elastic slope
        \cite{okorokov-arxiv-1501.01142} as well as in jet physics
        \cite{Okorokov-IJMPA-27-1250037-2012} and femtoscopy
        \cite{Okorokov-AHEP-2015-790646-2015}.} because the $\rho^{pp}$
    was measured at the same experimental conditions (detector,
    kinematic parameters etc.). The weighted average
    \cite{PTEP-2020-083C01-2020} can be used as estimation for
    $\rho^{pp}$ at $\sqrt{s}=13$ TeV: $\left.\langle
    \rho^{pp}\rangle\right|_{\sqrt{s}=13\,\scriptsize{\mbox{TeV}}}=0.095
    \pm 0.007$. Therefore, two versions of the DB20+ are considered
    here, namely, the database with one value $\left.\langle
    \rho^{pp}\rangle\right|_{\sqrt{s}=13\,\scriptsize{\mbox{TeV}}}$ is
    the DB20$_{1}$+ while the database contained both results
    ($\rho_{1}^{pp}$, $\rho_{2}^{pp}$) at $\sqrt{s}=13$ TeV is denoted
    as DB20$_{2}$+. Table \ref{tab:3} summarizes the main features of
    the databases of experimental results used in the present work for
    the set of the scattering parameters $\mathcal{G}_{1}$.

    Within the present work main goals, the values for each term
    $\{\mathcal{G}_{2}^{i}\}_{i=1}^{3}$ are calculated with the help
    of relations (\ref{eq:2.5}), (\ref{eq:asymp_1b}) and measured
    values of $\sigma_{\scriptsize\mbox{tot}}^{pp}$ and
    $\sigma_{\scriptsize\mbox{tot}}^{\bar{p}p}$ for completeness of
    analysis. Below, these ensembles of calculated values are called
    as data samples for $\mathcal{G}_{2}$ in similar with the set
    $\mathcal{G}_{1}$ just in the sense that value of each term
    $\{\mathcal{G}_{2}^{i}\}_{i=1}^{3}$ at certain $s$ is only defined
    by experimental results for $xp$ cross sections. The data sample
    for each term of $\mathcal{G}_{2}$ is based on the corresponding
    subensembles for $\sigma_{\scriptsize\mbox{tot}}^{pp}$ and
    $\sigma_{\scriptsize\mbox{tot}}^{\bar{p}p}$ at identical or, at
    least, close energies\footnote{The energy values for the measured
        $\sigma_{\scriptsize\mbox{tot}}^{pp}$ and
        $\sigma_{\scriptsize\mbox{tot}}^{\bar{p}p}$ are considered close
        if $P_{\Delta} \equiv |P_{1}-P_{2}| \leq 0.02$ GeV or $P_{1}$ and
        $P_{2}$ coincide within errors, where $P_{i}$ is the laboratory
        momentum for the term $\mathcal{G}_{1}^{i}$, $i=1,2$. On can note
        that the first condition for the closeness of $P_{1}$ and $P_{2}$
        was previously used in \cite{PAN-82-134-2019}. The relative
        fraction of the pairs ($\sigma_{\scriptsize\mbox{tot}}^{pp}$,
        $\sigma_{\scriptsize\mbox{tot}}^{\bar{p}p}$) with finite
        $P_{\Delta}$ is 64.8\% and almost all of such measurements are at
        $\sqrt{s} < 3.63$ GeV. In these cases the average momentum
        $\langle P\rangle$ is assigned for the initial energy for
        corresponding estimation of the each term $\mathcal{G}_{2}^{i}$,
        $i=1-3$ and $\langle P\rangle$ is calculated as simple average
        $\langle P\rangle = 0.5(P_{1}+P_{2})$ with $\Delta\langle P\rangle
        = 0.5|P_{1}-P_{2}|$ if $\exists\,i: \Delta P_{i}=0$ or $\langle
        P\rangle$ and its uncertainty is estimated with the help of the
        weighted average technique \cite{PTEP-2020-083C01-2020} on the
        contrary case.}. As seen from Table \ref{tab:3}, the subensembles
    for $\sigma_{\scriptsize\mbox{tot}}^{pp}$ and
    $\sigma_{\scriptsize\mbox{tot}}^{\bar{p}p}$ from
    \cite{PTEP-2020-083C01-2020} are only used for calculation of the
    each term of $\mathcal{G}_{2}$ and, consequently, there is one
    database for $\mathcal{G}_{2}$ corresponded to the DB20 for
    $\mathcal{G}_{1}$. As expected from the definitions (\ref{eq:2.5})
    and (\ref{eq:asymp_1b}), the energy range covered by experiments
    is identical for all terms of $\mathcal{G}_{2}$. Detailed analysis
    shown that this range is limited to $\sqrt{s} < 0.5$ TeV with wide
    gap between the highest ISR energy $\sqrt{s} \approx 0.06$ TeV and
    the high energy boundary $\sqrt{s}=(0.47 \pm 0.08)$ TeV for the
    energy domain under discussion.

    There is no prediction for the asymptotic energy $\sqrt{s_{a}}$
    from the first principles of QCD (for instance) as well as the
    search for the onset of the asymptotic energy domain remains a
    non-trivial task. Then, there is no consensus for the beginning of
    the so-called asymptotia. For example, the definition of the
    asymptotic regime may be done by using the first change of sign of
    the curvature parameter $C$ in the impact parameter representation
    \cite{M.M.Block.R.N.Cahn.Phys.Lett.B149.245.1984}. For the
    Chou--Yang model \cite{T.T.Chou.C.N.Yang.Phys.Rev.170.1591.1968},
    for instance, the asymptotia begins at $\sqrt{s_a}\approx 2$ TeV,
    where $C$ changes its sign. However, recent studies show
    model-dependent estimations for $s_{a}$, and these values lies in
    a very wide energy range. From the experimental point of view, it
    seems the most optimistic estimation for $\sqrt{s_{a}}$ is
    $\mathcal{O}$(100 TeV) in order of magnitude, and it was
    qualitatively obtained from the study of the functional behavior
    of $\sigma_{\footnotesize\mbox{tot}}^{pp}(s)$ within AQFT and CGC
    approach at ultra-high energies \cite{PAN-81-508-2018}. This
    result agrees with the conclusion from the Regge--eikonal model
    for $\sigma_{\footnotesize\mbox{tot}}^{pp}(s)$ and forward slope
    for $pp$ interaction \cite{IJMPA-33-1850077-2018}.

    \begin{table*}[!h]
        \caption{\label{tab:3}Databases for the set of global scattering
            parameters $\mathcal{G}_{1}$.}\label{tab:3}
        \begin{center}
            \begin{tabular}{ccccc}
                \hline \multicolumn{1}{c}{Database} & \multicolumn{4}{c}{Parameter
                    from the set $\mathcal{G}_{1}$} \\\cline{2-5}\rule{0pt}{10pt}
                & $\sigma_{\scriptsize\mbox{tot}}^{pp}$& $\sigma_{\scriptsize\mbox{tot}}^{\bar{p}p}$ & $\rho^{pp}$& $\rho^{\bar{p}p}$
                \rule{0pt}{10pt}\\
                \hline
                DB20        & ~~~\cite{PTEP-2020-083C01-2020} & ~~~\cite{PTEP-2020-083C01-2020}  & ~~~\cite{PTEP-2020-083C01-2020} & ~~~\cite{PTEP-2020-083C01-2020} \rule{0pt}{10pt}\\
                DB20$_{1}$+ & ~~~\cite{PTEP-2020-083C01-2020,PLB-808-135663-2020}
                & ~~~--//-- & ~~~\cite{PTEP-2020-083C01-2020} and $\left.\langle
                \rho^{pp}\rangle\right|_{\sqrt{s}=13\,\scriptsize{\mbox{TeV}}}$ & ~~~--//-- \rule{0pt}{10pt}\\
                DB20$_{2}$+ & ~~~--//-- & ~~~--//-- & ~~~\cite{EPJC-79-785-2019,PTEP-2020-083C01-2020} & ~~~--//-- \rule{0pt}{10pt}\\
                \hline
            \end{tabular}
        \end{center}
    \end{table*}

    Detailed analysis
    of the ratio of the elastic--to--total cross sections for $pp$ and
    $\bar{p}p$ collisions \cite{PAN-82-134-2019} as well as the
    approach of the partonic disks \cite{PU-185-963-2015} allow only
    the indication $\sqrt{s_{a}} \sim 5-10$ PeV. Consideration of some
    other signatures of the ''truly asymptotic regime" within
    Regge--eikonal model \cite{IJMPA-33-1850077-2018} results in much
    more conservative estimation for $s_{a}$, in particular, the onset
    of the asymptotic regime can be expected in grand unified theory
    (GUT) energy domain in order of magnitude, i.e. $\sqrt{s_{a}}
    \gtrsim 10^{12}-10^{13}$ GeV.

    As seen, there are only 1 -- 2 measurements for
    $\sigma_{\scriptsize\mbox{tot}}^{pp}$ in ultra-high energy cosmic
    rays even for lowest estimation for $\sqrt{s_{a}}$. Therefore, a
    phenomenological approximation will be \emph{a priori} at
    collision energies smaller than the possible onset of the
    asymptotic region. Consequently, the request for the validity of
    the Pomeranchuk theorem seems redundant for the energy range under
    fit, and one can consider the hypothesis (ii) as well as (iii) for
    the functional forms of the terms of $\mathcal{G}$ within the
    fitting procedure. As previously
    \cite{IJMPA-A25-5333-2010,IJMPA-32-1750175-2017}, the
    parameterizations shown in Table \ref{tab:1} and \ref{tab:2} for
    hypotheses (ii), (iii) will be applied for approximation of the
    energy dependence of different terms of $\mathcal{G}$ only for $s
    \geq s_{\scriptsize\mbox{min}}$, where $s_{\scriptsize\mbox{min}}$
    is some empirical low boundary. During the analysis procedure, the
    $s_{\scriptsize\mbox{min}}$ value will be decreased as much as
    possible in order to describe the wider energy domain with
    statistically reasonable fit quality.

    \section{Results of Simultaneous Fits}\label{sec:exp-1}

    The Section contains the detailed description of the results of
    simultaneous fits for the sets $\mathcal{G}_{i}$, $i=1, 2$ and
    corresponding discussion.

    \subsection{Simultaneous Fits for the Set $\mathcal{G}_{1}$}\label{subsec:exp-1-1}

    The energy dependence of terms of $\mathcal{G}_{1}$ is
    approximated at $\sqrt{s_{\scriptsize\mbox{min}}}=0.03, 0.04,
    0.05, 0.06, 0.1, 0.5, 1, 5$ and 10 TeV by the corresponding
    formulas from Table \ref{tab:2} within hypothesis (ii) and (iii).
    At lowest $s_{\scriptsize\mbox{min}}$ considered here the fit
    quality $\chi^{2}/\mbox{n.d.f.} \approx 26$ with fast decrease at
    growth of the low boundary of fitted range for hypotheses (ii) and
    (iii). For both hypotheses considered the statistically reasonable
    fit qualities are only observed for
    $\sqrt{s_{\scriptsize\mbox{min}}} \geq 0.06$ TeV. The present
    approach allows the reasonable description of $\mathcal{G}_{1}$
    within narrower energy ranges than that AQFT equations
    \cite{IJMPA-A25-5333-2010,IJMPA-32-1750175-2017} with shifting
    $s_{\scriptsize\mbox{min}}$ towards larger values. In general,
    this result is expected because the asymptotic behavior of the
    total cross sections and $\rho$--parameter is studied here for
    $pp$ and $\bar{p}p$ collisions. Thus the discussion below is
    focused on the results for $\sqrt{s_{\scriptsize\mbox{min}}} \geq
    0.06$ TeV.

    \begin{table*}
        \caption{\label{tab:4}Parameters for simultaneous fitting of $\mathcal{G}_{1}(s)$ within various hypotheses at different stages of DB: DB20 (first line) and DB20$_{1}$+ (second line).}
        \begin{center}
                \begin{tabular}{lcccccccccc}
                    \hline \multicolumn{1}{l}{$\sqrt{s_{\scriptsize{\mbox{min}}}}$,} &
                    \multicolumn{4}{c}{hypothesis (ii)} & \multicolumn{3}{c}{} &
                    \multicolumn{3}{c}{hypothesis (iii)} \rule{0pt}{10pt}\\
                    \cline{2-11}
                    TeV & $c_{p}$, mbarn & $c_{\bar{p}}$, mbarn & $a_{2}$, mbarn & $\chi^{2}/\mbox{n.d.f.}$ & & & & $c$, mbarn & $a_{1}$, mbarn & $\chi^{2}/\mbox{n.d.f.}$ \rule{0pt}{10pt}\\
                    \hline
                    0.06 & $8.484 \pm 0.029$ & $8.12 \pm 0.04$ & $(-9.0 \pm 0.9) \times 10^{3}$ & $97.7/45$ & & & & $8.342 \pm 0.024$ & $(-1.9 \pm 0.6) \times 10^{3}$ & $170/46$\rule{0pt}{10pt}\\
                    & $8.468 \pm 0.029$ & $8.12 \pm 0.04$ & $(-8.5 \pm 0.9) \times 10^{3}$ & $108/47$ & & & & $8.339 \pm 0.024$ & $(-1.8 \pm 0.6) \times 10^{3}$ & $174/48$ \\
                    0.1 & $8.35 \pm 0.04$ & $7.91 \pm 0.06$ & $(-1.0 \pm 0.4) \times 10^{5}$ & $39.6/34$ & & & & $8.31 \pm 0.04$ & $(2.8 \pm 1.5) \times 10^{5}$ & $124/35$ \\
                    & $8.35 \pm 0.04$ & $7.91 \pm 0.06$ & $(-1.2 \pm 0.5) \times 10^{5}$ & $43.7/36$ & & & & $8.28 \pm 0.04$ & $(2.2 \pm 0.9) \times 10^{4}$ & $134/37$ \\
                    0.5 & $8.38 \pm 0.05$ & $8.00 \pm 0.11$ & $(3.1 \pm 1.4) \times 10^{5}$ & $37.1/30$ & & & & $8.48 \pm 0.05$ & $(2.04 \pm 0.25) \times 10^{6}$ & $62.3/31$ \\
                    & $8.39 \pm 0.05$ & $8.01 \pm 0.11$ & $(3.5 \pm 1.2) \times 10^{5}$ & $41.0/31$ & & & & $8.48 \pm 0.05$ & $(2.0 \pm 0.4) \times 10^{6}$ & $65.2/32$   \\
                    1 & $8.28 \pm 0.17$ & $7.6 \pm 0.6$ & $(-1.9 \pm 0.5) \times 10^{7}$ & $33.0/23$ & & & & $8.54 \pm 0.05$ & $(1.8 \pm 0.4) \times 10^{7}$ & $34.9/24$ \\
                    & $8.4 \pm 0.3$   & $8.0 \pm 1.1$ & $(-2.4 \pm 0.8) \times 10^{6}$ & $37.3/24$ & & & & $8.54 \pm 0.05$ & $(1.9 \pm 0.4) \times 10^{7}$ & $37.9/25$ \\
                    5 & $9.3 \pm 0.4$   & $9.9 \pm 2.6$  & $(-3.1 \pm 0.6) \times 10^{8}$ & $20.3/17$ & & & & $9.03 \pm 0.21$ & $(-3.1 \pm 0.8) \times 10^{8}$ & $20.5/18$  \\
                    & $9.35 \pm 0.15$ & $10.0 \pm 1.9$ & $(-3.1 \pm 0.6) \times 10^{8}$ & $21.4/18$ & & & & $9.06 \pm 0.22$ & $(-3.3 \pm 0.9) \times 10^{8}$ & $22.8/19$  \\
                    10 & $9.4 \pm 1.7$ & $10.1 \pm 2.7$ & $(6.5 \pm 0.5) \times 10^{7}$  & $1.08/6$ & & & & $9.1 \pm 1.0$ & $(6.5 \pm 0.7) \times 10^{7}$ & $1.08/7$ \\
                    & $9.8 \pm 0.5$ & $10.0 \pm 2.6$ & $(-1.3 \pm 0.9) \times 10^{9}$ & $2.16/7$ & & & & $9.9 \pm 0.6$ & $(-1.7 \pm 0.8) \times 10^{9}$ & $2.26/8$ \\
                    \hline
                \end{tabular}
        \end{center}
    \end{table*}

    As seen from Table \ref{tab:3}, the data bases considered here
    differ from each other very slightly, more precisely, the maximum
    difference on 3 points is between DB20 and DB20$_{2}$+.
    Furthermore, all of the experimental results which are addition
    with respect to the DB20 agree well with the general trends in the
    energy dependence of the corresponding observable. Therefore, in
    accordance with the hypothesis confirmed in
    \cite{IJMPA-32-1750175-2017} one can expect the negligible affect
    of the addition points on the values of fit parameters for various
    data bases at fixed $s_{\scriptsize{\mbox{min}}}$. Detailed
    analysis fully confirms the correctness of this suggestion for the
    results of simultaneous fits for $\mathcal{G}_{1}$ with
    data bases DB20$_{1}$+ and DB20$_{2}$+ differ from each other only
    one point (Table \ref{tab:3}). The values of all fit parameters
    agrees within errors for DB20$_{1}$+ and DB20$_{2}$+ for each
    $s_{\scriptsize{\mbox{min}}}$ and hypotheses (ii), (iii)
    considered here. Moreover, the identity is observed between
    numerical values of fit parameters and their uncertainties for
    DB20$_{1}$+ and values for corresponding quantity for DB20$_{2}$+
    for the noticeable part of low boundaries
    $s_{\scriptsize{\mbox{min}}}$, especially within the hypothesis
    (iii). The approximation quality $\chi^{2}/\mbox{n.d.f.}$ is very
    close for the simultaneous fits of DB20$_{1}$+ and DB20$_{2}$+
    with subtle improvement for the last case at any fixed
    $s_{\scriptsize{\mbox{min}}}$. All of these allows us to consider
    below the fit results obtained for DB20 and DB20$_{1}$+.

\begin{figure}
\centering
\includegraphics[width=17.0cm,height=17.0cm]{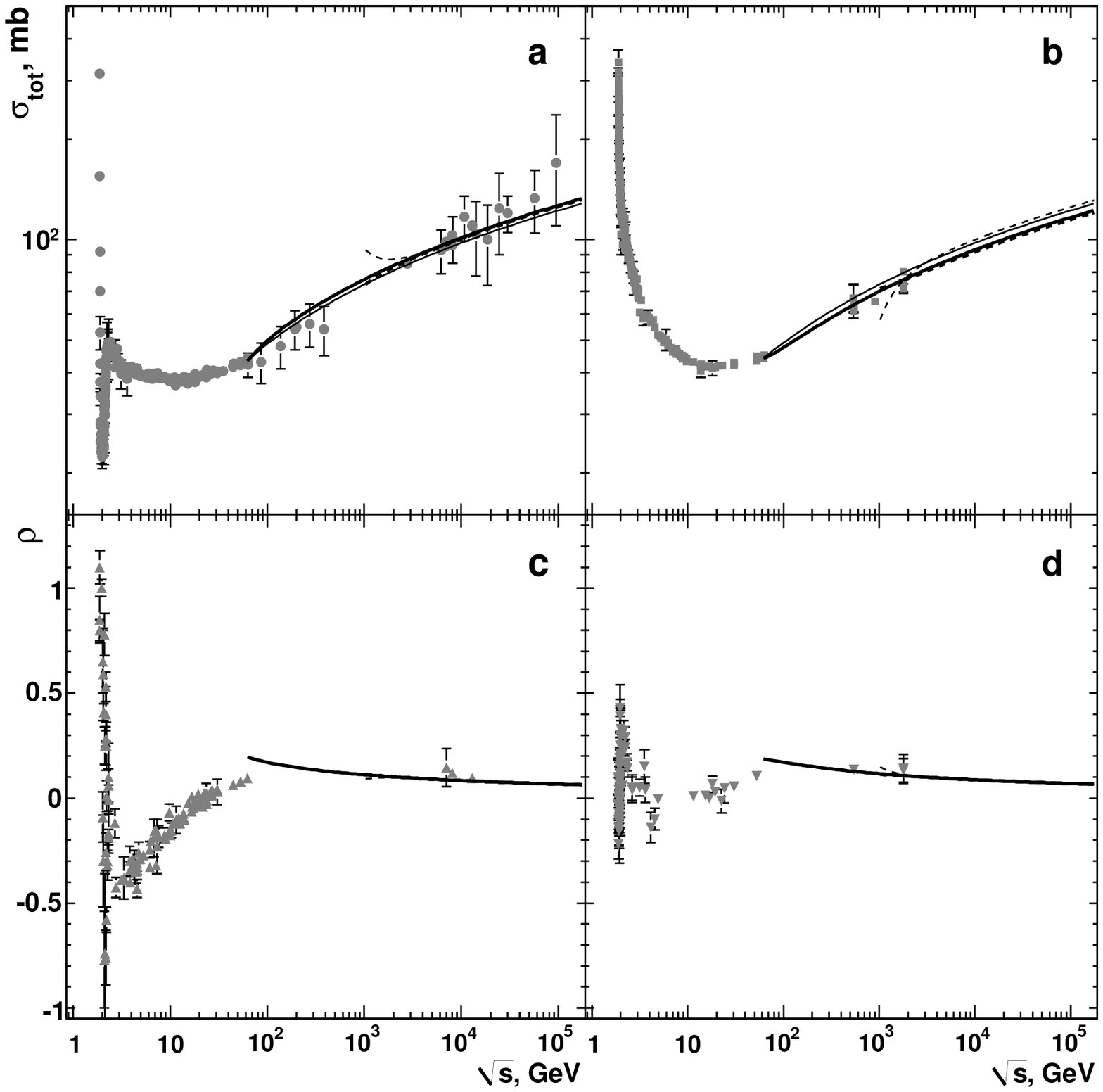}
\caption{The energy dependence of the measurements for the terms of $\mathcal{G}_{1}$ and results of simultaneous fits of all these terms. Experimental results are from the DB20$_{1}$+. The solid lines
correspond to the fit at $\sqrt{s_{\scriptsize\mbox{min}}}=0.06$ TeV, the dashed lines are
at $\sqrt{s_{\scriptsize\mbox{min}}}=1$ TeV. The thick lines show the fit curves for the hypothesis (ii) and
the other lines correspond to the hypothesis (iii).} \label{fig:1}
\end{figure}

    The numerical results of the simultaneous fits of
    $\mathcal{G}_{1}$ in various energy ranges are shown in Tables
    \ref{tab:4} for hypothesis (ii) and (iii). At fixed
    $s_{\scriptsize{\mbox{min}}}$ the first line is for DB20 as well
    as the second line shows results for DB20$_{1}$+. Experimental
    data from DB20$_{1}$+ for the terms $\mathcal{G}_{1}^{i}$, $i=1-4$
    together with fit curves are shown in Fig. \ref{fig:1} for
    $\sqrt{s_{\scriptsize\mbox{min}}}=0.06$ TeV (solid lines) and at
    $\sqrt{s_{\scriptsize\mbox{min}}}=1$ TeV (dashed lines). The thick
    curves are obtained within the hypothesis (ii) while the two
    remaining lines correspond to the hypothesis (iii).

    The use of the multi-TeV values of
    $\sqrt{s_{\scriptsize\mbox{min}}}$ and the available experimental
    data within certain data base stipulates that the approximation
    procedure consequently transits from the simultaneous fit of the
    full set $\mathcal{G}_{1}$ to the simultaneous fit of the $pp$
    observables $\{\sigma_{\scriptsize{\mbox{tot}}}^{pp},\rho^{pp}\}$
    only at $\sqrt{s_{\scriptsize\mbox{min}}}=5$ TeV and even to the
    individual fit of the $\sigma_{\scriptsize{\mbox{tot}}}^{pp}$ at
    the highest $\sqrt{s_{\scriptsize\mbox{min}}}=10$ TeV in the case
    of the DB20. In the last case, strictly speaking, hypothesis (ii)
    reduces to the hypothesis (iii) because of available experimental
    data allow the fix only the combination $(3c_{p}-c_{\bar{p}})$.
    There is no fitting function for which there would be, at least,
    one experimental point at $\sqrt{s} \geq 10$ TeV and, at the same
    time, this function would contain parameter $c_{p}$ or
    $c_{\bar{p}}$ in the separate term. Moreover, one expects the
    smooth joining for the energy-dependent $\bar{p}p$ global
    observables, $\sigma_{\scriptsize\mbox{tot}}^{\bar{p}p}(s)$ and
    $\rho^{\bar{p}p}(s)$, in the experimentally measured range as well
    in the domain $\sqrt{s} \geq 5\,(10)$ TeV described by the curve
    calculated from the fit results at
    $\sqrt{s_{\scriptsize\mbox{min}}}=5\,(10)$ TeV. This expectation
    is established by the absence of the any signature for new physics
    beyond the Standard Model (SM) which could result in the sharp
    changing of the energy dependence of any global scattering
    parameter in $pp/\bar{p}p$ collisions.

    There are only few points for $\rho^{xp}(s)$ in the TeV-energy
    domain and, moreover, the experimental values $\rho^{xp} \sim
    10^{-2}$ at $\sqrt{s} > 1$ TeV; the consideration aforementioned
    in Sec. \ref{sec:model} implies the smooth behavior of the curves
    for $\rho^{xp}(s)$ is dominated by the values of $c_{x}$ / $c$
    parameters within hypothesis (ii) / (iii).  Furthermore, changes
    in these curves are slow ($\propto \ln^{-1}\varepsilon$) at
    sufficiently high $s$. All of these evidences result in relatively
    robust behavior of the curve for $\rho^{\bar{p}p}(s)$ at
    $\sqrt{s_{\scriptsize\mbox{min}}}=5$, 10 TeV as well as a
    reasonable agreement between experimental value of
    $\left.\rho^{\bar{p}p}(s)\right|_{\sqrt{s} \approx
        2\,\scriptsize\mbox{TeV}}$ and analytic approximation in multi-TeV
    energy domain without any additional request for the smooth
    joining.
    Detailed analysis shows some influence of the
    request of the smooth joining for
    $\sigma_{\scriptsize\mbox{tot}}^{\bar{p}p}(s)$ on the fit results
    within hypothesis (ii) for DB20$_{1}$+ at
    $\sqrt{s_{\scriptsize\mbox{min}}}=10$ TeV. For this case the
    numerical values of the fit parameters obtained taking into
    account the need of the smooth joining for
    $\sigma_{\scriptsize\mbox{tot}}^{\bar{p}p}(s)$ are shown in Table
    \ref{tab:4} while the results without additional request are
    following: $c_{p}=(10.5 \pm 0.8)$ mbarn, $c_{\bar{p}}=(13.2 \pm
    2.8)$ mbarn, $a_{2}=(6.0 \pm 0.6) \times 10^{7}$ mbarn,
    $\chi^{2}/\mbox{n.d.f.}=1.08/7$.

    As seen the request of the smooth joining for
    $\sigma_{\scriptsize\mbox{tot}}^{\bar{p}p}(s)$ influences on the
    value of $a_{2}$ parameter and fit quality. If one releases the
    need under consideration, then one provides the agreement between
    values of $a_{2}$ for various data bases within hypothesis (ii) at
    almost identical $\chi^{2}/\mbox{n.d.f.}$. Such exception results
    in the expected discrepancy between values of $a_{2}$ and
    $\chi^{2}/\mbox{n.d.f.}$ obtained by simultaneous fits within
    various hypotheses for data base DB20$_{1}$+. One can note the
    change in the fit quality for the last two cases is noticeable but
    not at a first glance. It is important to point out that
    $\chi^{2}/\mbox{n.d.f.}$ remains statistically acceptable for
    simultaneous fit of DB20$_{1}$+ within hypothesis (ii) at highest
    $s_{\scriptsize\mbox{min}}$, independently on additional request
    of the smooth joining for
    $\sigma_{\scriptsize\mbox{tot}}^{\bar{p}p}(s)$.

    As observed from Table \ref{tab:4}, the hypothesis (ii) allows the
    simultaneous approximation of all terms of $\mathcal{G}_{1}$ with
    reasonable quality at $\sqrt{s_{\scriptsize\mbox{min}}} \geq 0.06$
    TeV and with statistically acceptable one at
    $\sqrt{s_{\scriptsize\mbox{min}}} \geq 0.1$ TeV for any data bases
    considered. The following relation $c_{p} > c_{\bar{p}}$ is valid
    for the simultaneous fit versions within hypothesis (ii) up to
    $\sqrt{s_{\scriptsize\mbox{min}}}=0.5$ TeV. On the other hand, the
    values of $c_{p}$ and $c_{\bar{p}}$ coincide within errors for
    approximations at $\sqrt{s_{\scriptsize\mbox{min}}} \geq 1$ TeV.
    These statements are valid for both data bases DB20 and
    DB20$_{1}$+. It is important to note that for the hypothesis
    (iii), in contrast with (ii), a reasonable values of
    $\chi^{2}/\mbox{n.d.f.}$ can be obtained only at
    $\sqrt{s_{\scriptsize\mbox{min}}} \geq 0.5$ TeV and statistically
    acceptable $\chi^{2}/\mbox{n.d.f.}$ -- only at
    $\sqrt{s_{\scriptsize\mbox{min}}} \geq 1$ TeV. This feature is in
    full agreement with Table \ref{tab:4}: as consequence of the
    considered relations between $c_{p}$ and $c_{\bar{p}}$ at various
    $s_{\scriptsize\mbox{min}}$, it can be expected that reducing the
    parameters $c_{p}$, $c_{\bar{p}}$ to $c$ would be possible
    starting, at least, at similar collision
    energies\footnote{Strictly speaking, the possible reduction of the
        number of fit parameters does not mean the transition from the
        hypothesis (ii) to the (iii) one because the corresponding fit
        parameters obtained within these hypotheses disagree for DB20 at
        $\sqrt{s_{\scriptsize\mbox{min}}}=1$ TeV and similar statement is
        for $a_{1}$ and $a_{2}$ parameters in the case of DB20$_{1}$+.}.
    Thus, hypothesis (iii) reasonably describes the experimental data
    for $\mathcal{G}_{1}$ at substantially higher $s$ than the
    hypothesis (ii) due to ''extremely" asymptotic nature of the
    corresponding relations in Table \ref{tab:2}. The hypotheses (ii)
    and (iii) qualitatively describe the energy-dependent behavior of
    $\sigma_{\scriptsize\mbox{tot}}^{xp}$ at
    $\sqrt{s_{\scriptsize\mbox{min}}}=0.06$ TeV already (Fig.
    \ref{fig:1}a, b). However, they give overestimates for the
    $\rho^{xp}$ at $\sqrt{s} < 0.1$ TeV (Fig. \ref{fig:1}c, d) which
    allows the qualitative assumption that discrepancy between
    experimental values of $\rho^{xp}$ and smooth curves within
    hypotheses (ii) and (iii) at $\sqrt{s} < 0.1$ TeV is the main
    reason for the large values of $\chi^{2}/\mbox{n.d.f.}$ for
    simultaneous fits at small values of low boundary of the fitted
    range considered here ($\sqrt{s_{\scriptsize\mbox{min}}}=0.03,
    0.04$ and 0.05 TeV).

    Comparative analysis of the fit parameters obtained for data bases
    DB20 and DB20$_{1}$+ and shown in Table \ref{tab:4} results in the
    following conclusions: (a) close values of
    $\chi^{2}/\mbox{n.d.f.}$ are for both data bases at any fixed
    $s_{\scriptsize\mbox{min}}$; (b) values of $c_{p}$, $c_{\bar{p}}$
    within hypothesis (ii) agrees within errors for various data bases
    at corresponded $s_{\scriptsize\mbox{min}}$ as well as values  of
    $c$ in the case of the hypothesis (iii); (c) mostly the last
    statement is also valid for parameter $a_{2} / a_{1}$ for the
    hypothesis (ii) / (iii).

    In general, the versions (ii) and (iii) of the asymptotic model
    suggested within the present work describe the experimental data
    for the set $\mathcal{G}_{1}$ at higher energies than AQFT
    \cite{IJMPA-A25-5333-2010,IJMPA-32-1750175-2017}, especially the
    hypothesis (iii). The comparison between the asymptotic model and
    AQFT is possible only at $\sqrt{s_{\scriptsize\mbox{min}}}=0.06$
    TeV, and fit quality is noticeably worse in the first case than
    that for AQFT. Such a relationship between the phenomenological
    models is expected, since the approach considered here is \emph{a
        priori} asymptotic, i.e. as reasonably expect the model based on
    the suggestions allowed for the asymptotic energy domain will
    describe of the experimental data well at higher, strictly
    speaking, asymptotically high energies.

    \begin{table*}
        \caption{\label{tab:5}Parameters for simultaneous fitting of
            $\mathcal{G}_{1}(s)$ within various hypotheses at different stages
            of the accelerator subsample of DB:
            DB$_{\scriptsize{\mbox{ac}}}$20 (first line) and
            DB$_{\scriptsize{\mbox{ac}}}$20$_{1}$+ (second line).}
        \begin{center}
                \begin{tabular}{lcccccccccc}
                    \hline \multicolumn{1}{l}{$\sqrt{s_{\scriptsize{\mbox{min}}}}$,} &
                    \multicolumn{4}{c}{hypothesis (ii)} & \multicolumn{3}{c}{} &
                    \multicolumn{3}{c}{hypothesis (iii)} \rule{0pt}{10pt}\\
                    \cline{2-11}
                    TeV & $c_{p}$, mbarn & $c_{\bar{p}}$, mbarn & $a_{2}$, mbarn & $\chi^{2}/\mbox{n.d.f.}$ & & & & $c$, mbarn & $a_{1}$, mbarn & $\chi^{2}/\mbox{n.d.f.}$ \rule{0pt}{10pt}\\
                    \hline
                    0.06 & $8.49 \pm 0.03$ & $8.12 \pm 0.04$ & $(-9.1 \pm 0.9) \times 10^{3}$ & $92.1/31$ & & & & $8.343 \pm 0.024$ & $(-1.9 \pm 0.6) \times 10^{3}$ & $165/32$\rule{0pt}{10pt}\\
                    & $8.471 \pm 0.029$ & $8.12 \pm 0.04$ & $(-8.6 \pm 0.9) \times 10^{3}$ & $103/33$ & & & & $8.339 \pm 0.024$ & $(-1.8 \pm 0.6) \times 10^{3}$ & $169/34$ \\
                    0.1 &  & -- &  & & & & & & -- & \\
                    & $8.35 \pm 0.04$ & $7.91 \pm 0.06$ & $(-1.2 \pm 0.6) \times 10^{5}$ & $40.2/23$ & & & & $8.28 \pm 0.04$ & $(4.4 \pm 0.8) \times 10^{4}$ & $129/24$ \\
                    0.5 & $8.38 \pm 0.05$ & $8.00 \pm 0.11$ & $(3.1 \pm 1.4) \times 10^{5}$ & $35.0/21$ & & & & $8.48 \pm 0.05$ & $(2.03 \pm 0.25) \times 10^{6}$ & $59.9/22$ \\
                    & $8.38 \pm 0.05$ & $8.01 \pm 0.11$ & $(3.5 \pm 0.9) \times 10^{5}$ & $39.0/22$ & & & & $8.48 \pm 0.05$ & $(2.03 \pm 0.25) \times 10^{6}$ & $62.9/23$   \\
                    1 & $8.29 \pm 0.16$ & $7.7 \pm 0.5$ & $(-1.7 \pm 0.8) \times 10^{7}$ & $31.0/14$ & & & & $8.54 \pm 0.05$ & $(1.9 \pm 0.4) \times 10^{7}$ & $32.8/15$ \\
                    & $8.39 \pm 0.05$   & $8.02 \pm 0.11$ & $(-2.4 \pm 0.5) \times 10^{6}$ & $35.3/15$ & & & & $8.54 \pm 0.05$ & $(1.9 \pm 0.4) \times 10^{7}$ & $35.8/16$ \\
                    5 & $9.34 \pm 0.16$   & $10.0 \pm 2.7$  & $(-2.9 \pm 0.7) \times 10^{8}$ & $19.0/8$ & & & & $9.01 \pm 0.23$ & $(-3.0 \pm 0.9) \times 10^{8}$ & $19.2/9$  \\
                    & $9.33 \pm 0.16$ & $10.0 \pm 1.9$ & $(-2.9 \pm 0.8) \times 10^{8}$ & $20.1/9$ & & & & $9.04 \pm 0.23$ & $(-3.2 \pm 0.7) \times 10^{8}$ & $21.5/10$  \\
                    10 & & -- & & & & & & & -- & \\
                    & & -- & & & & & & $10.5 \pm 0.7$ & $(-2.7 \pm 0.7) \times 10^{9}$ & $0.05/1$ \\
                    \hline
                \end{tabular}
        \end{center}
    \end{table*}

    The ensemble of experimental results that includes only
    accelerator data is also considered for each data base DB20,
    DB20$_{1}$+ and DB20$_{2}$+. These ensembles are denoted as
    DB$_{\scriptsize{\mbox{ac}}}$20,
    DB$_{\scriptsize{\mbox{ac}}}$20$_{1}$+ and
    DB$_{\scriptsize{\mbox{ac}}}$20$_{2}$+. The energy dependence of
    terms of  $\mathcal{G}_{1}$ is approximated at
    $\sqrt{s_{\scriptsize\mbox{min}}}=0.03, 0.04, 0.05, 0.06, 0.1,
    0.5, 1, 5$ and 10 TeV by the corresponding formulas from Table
    \ref{tab:2} within hypothesis (ii) and (iii) for each ensemble
    DB$_{\scriptsize{\mbox{ac}}}$20,
    DB$_{\scriptsize{\mbox{ac}}}$20$_{1}$+ and
    DB$_{\scriptsize{\mbox{ac}}}$20$_{2}$+ as well as for the full
    data bases above. The statement made for DB20, DB20$_{1}$+ and
    DB20$_{2}$+ with regards of fit qualities is valid for this case.
    Therefore the discussion below for samples
    DB$_{\scriptsize{\mbox{ac}}}$20,
    DB$_{\scriptsize{\mbox{ac}}}$20$_{1}$+ and
    DB$_{\scriptsize{\mbox{ac}}}$20$_{2}$+  is focused on the results
    for $\sqrt{s_{\scriptsize\mbox{min}}} \geq 0.06$ TeV.

    The values of all fit parameters agree quite well within errors
    for DB$_{\scriptsize{\mbox{ac}}}$20$_{1}$+ and
    DB$_{\scriptsize{\mbox{ac}}}$20$_{2}$+ for each
    $s_{\scriptsize{\mbox{min}}}$ and hypotheses (ii), (iii)
    considered here. The approximation quality
    $\chi^{2}/\mbox{n.d.f.}$ is very close for the simultaneous fits
    of accelerator data ensembles
    DB$_{\scriptsize{\mbox{ac}}}$20$_{1}$+ and
    DB$_{\scriptsize{\mbox{ac}}}$20$_{2}$+ with subtle improvement for
    the last case at any fixed $s_{\scriptsize{\mbox{min}}}$. These
    statements are valid with exception of the highest
    $\sqrt{s_{\scriptsize\mbox{min}}}=10$ TeV analyzed below more
    detailed. As well as for the full data bases all of these allows
    us to consider below the fit results for
    DB$_{\scriptsize{\mbox{ac}}}$20 and
    DB$_{\scriptsize{\mbox{ac}}}$20$_{1}$+.

\begin{figure}
\centering
\includegraphics[width=17.0cm,height=17.0cm]{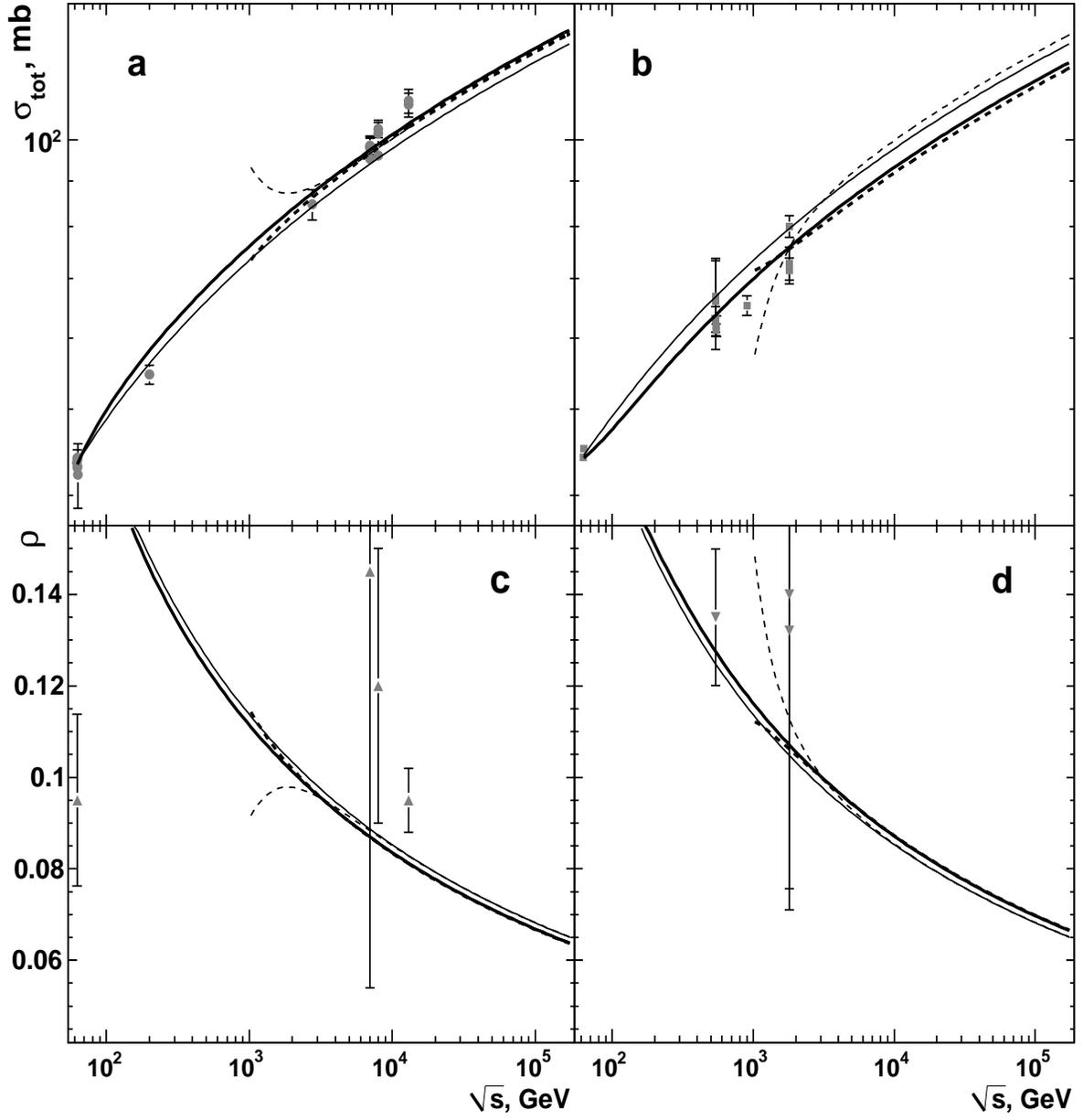}
\caption{The comparison of the data from accelerator experiments
for the terms of $\mathcal{G}_{1}$ and approximating
curves from simultaneous fits of all these terms at $\sqrt{s} \geq
0.06$ TeV. Experimental results are from the subsample
DB$_{\scriptsize\mbox{ac}}$20$_{1}$+. The notations for the curves
and experimental data base are identical to that in Fig.
\ref{fig:1}.} \label{fig:2}
\end{figure}

    The numerical results of the simultaneous fits of
    $\mathcal{G}_{1}$ in various energy ranges are shown in Tables
    \ref{tab:5} for hypothesis (ii) and (iii) taking into account the
    accelerator data only. At fixed $s_{\scriptsize{\mbox{min}}}$ the
    first line is for DB$_{\scriptsize{\mbox{ac}}}$20 and second line
    shows results for DB$_{\scriptsize{\mbox{ac}}}$20$_{1}$+.
    Experimental data from DB$_{\scriptsize{\mbox{ac}}}$20$_{1}$+ for
    the terms $\mathcal{G}_{1}^{i}$, $i=1-4$ together with fit curves
    are shown in Fig. \ref{fig:2} for
    $\sqrt{s_{\scriptsize\mbox{min}}}=0.06$ TeV (solid lines) and at
    $\sqrt{s_{\scriptsize\mbox{min}}}=1$ TeV (dashed lines). The thick
    curves are obtained within the hypothesis (ii) and the two other
    lines correspond to the hypothesis (iii). The energy domain
    $\sqrt{s} \geq 0.06$ TeV is considered in Fig. \ref{fig:2} in
    difference with Fig. \ref{fig:1} in order to show fit curves
    clearer for different $s_{\scriptsize\mbox{min}}$ and hypotheses.

    Numerical values for the fit parameters agree within errors for
    Tables \ref{tab:4} and \ref{tab:5} at fixed
    $s_{\scriptsize\mbox{min}}$ and hypothesis. Therefore, most of
    conclusions made above for the full data bases DB20, DB20$_{1}$+
    and DB20$_{2}$+ are valid for the corresponded ensembles of the
    accelerator experimental results with the following features and
    interpretations. There are agreement, within errors, between the
    numerical values of the corresponded fit parameters for ensembles
    DB$_{\scriptsize{\mbox{ac}}}$20 and
    DB$_{\scriptsize{\mbox{ac}}}$20$_{1}$+ for certain hypothesis and
    fixed $s_{\scriptsize\mbox{min}}$. Furthermore, values are
    identical for free parameters at
    $\sqrt{s_{\scriptsize\mbox{min}}}=0.5$ and 1 TeV within hypothesis
    (iii). As consequence, the fit curves for various hypotheses are
    close for each other at any fixed $s_{\scriptsize\mbox{min}}$
    considered in Fig. \ref{fig:2} for each observable from the set
    $\mathcal{G}_{1}$.

    Table \ref{tab:5} shows that there is noticeable decreasing of the
    relative uncertainties for $c$ at
    $\sqrt{s_{\scriptsize\mbox{min}}}=1$ TeV, for $c_{\bar{p}}$ at
    $\sqrt{s_{\scriptsize\mbox{min}}} \geq 1$ TeV, for $a_{2}$ at
    $\sqrt{s_{\scriptsize\mbox{min}}}=0.5$ and 1 TeV at transition
    from the ensemble DB$_{\scriptsize{\mbox{ac}}}$20 to
    DB$_{\scriptsize{\mbox{ac}}}$20$_{1}$+ one for hypothesis (ii).
    The similar effect is absent for pairs of any other ensembles at
    certain hypothesis. The data sets for approximation are identical
    for $\sqrt{s_{\scriptsize\mbox{min}}}=0.1$ and 0.5 TeV in the case
    of the sample DB$_{\scriptsize{\mbox{ac}}}$20 wherefore the fit
    results are only shown in the last case in Table \ref{tab:5}. The
    fits are impossible at highest
    $\sqrt{s_{\scriptsize\mbox{min}}}=10$ TeV due to lack of the
    required number of data points with exception of the cases of
    DB$_{\scriptsize{\mbox{ac}}}$20$_{1}$+ for hypothesis (iii), shown
    in Table \ref{tab:5} and ensemble
    DB$_{\scriptsize{\mbox{ac}}}$20$_{2}$+ for both hypotheses under
    study. In the last case the following results are obtained with
    help of the simultaneous fit at highest
    $\sqrt{s_{\scriptsize\mbox{min}}}=10$ TeV: $c_{p}=(10.00 \pm
    0.13)$ mbarn, $c_{\bar{p}}=(10.0 \pm 1.7)$ mbarn, $a_{2}=(-1.8 \pm
    1.0) \times 10^{9}$ mbarn, $\chi^{2}/\mbox{n.d.f.}=0.88/1$ for the
    hypothesis (ii) and $c=(10.5 \pm 0.7)$ mbarn, $a_{1}=(-2.7 \pm
    0.7) \times 10^{9}$ mbarn, $\chi^{2}/\mbox{n.d.f.}=0.55/2$ --
    hypothesis (iii).

    These data confirm all conclusions made above regarding of the
    agreement between fitting results for
    DB$_{\scriptsize{\mbox{ac}}}$20$_{1}$+ (Table \ref{tab:5}) and
    DB$_{\scriptsize{\mbox{ac}}}$20$_{2}$+ in the case of the
    hypothesis (iii) and between the full data base DB20$_{1}$+,
    DB20$_{2}$+ and corresponding ensembles of accelerator
    experimental results for hypothesis (ii) at
    $\sqrt{s_{\scriptsize\mbox{min}}}=10$ TeV taking into account the
    closeness of the results of approximation established above for
    DB20$_{1}$+ and DB20$_{2}$+.

    One can note that the fitting results are obtained with the
    request for the smooth joining of the
    $\sigma_{\scriptsize\mbox{tot}}^{\bar{p}p}(s)$ and
    $\rho^{\bar{p}p}(s)$ in experimentally measured range and in the
    domain $\sqrt{s} \geq 5\,(10)$ TeV within hypothesis (ii) at
    $\sqrt{s_{\scriptsize\mbox{min}}}=5$ TeV for the ensembles
    DB$_{\scriptsize{\mbox{ac}}}$20,
    DB$_{\scriptsize{\mbox{ac}}}$20$_{1}$+ (Table \ref{tab:5}) and at
    $\sqrt{s_{\scriptsize\mbox{min}}}=5$ and 10 TeV for
    DB$_{\scriptsize{\mbox{ac}}}$20$_{2}$+. As seen from Tables
    \ref{tab:4} and \ref{tab:5}, the exception of the cosmic ray (CR)
    measurements result in the deterioration in the quality of the
    approximation for the ensembles DB$_{\scriptsize{\mbox{ac}}}$20
    and DB$_{\scriptsize{\mbox{ac}}}$20$_{1}$+ regarding of the
    corresponded full data bases\footnote{The case of the
        $\sqrt{s_{\scriptsize\mbox{min}}}=10$ TeV is special one because
        there only are 1 or 2 n.d.f. if any for
        DB$_{\scriptsize{\mbox{ac}}}$20$_{1}$+ and
        DB$_{\scriptsize{\mbox{ac}}}$20$_{2}$+ respectively. Thus the
        results obtained for so small n.d.f. can not be considered as
        representative.}. Only statistically reasonable values of
    $\chi^{2}/\mbox{n.d.f.}$ are obtained for ensembles of accelerator
    results for any $\sqrt{s_{\scriptsize\mbox{min}}} \geq 0.06$ TeV
    in difference with the full data bases. Consequently, hypotheses
    (ii) and (iii) allow the qualitative description of energy
    dependence of $\mathcal{G}_{1}$ only. Fig. \ref{fig:2}
    demonstrates the qualitative agreement of the curves from
    simultaneous fits of the terms of $\mathcal{G}_{1}$ with data
    points at any considered $s_{\scriptsize\mbox{min}}$. The main
    features of the behavior of these fit curves with respect to
    the experimental results are similar to those observed in Fig.
    \ref{fig:1} at corresponded $s_{\scriptsize\mbox{min}}$.

    In general, one can conclude the exclusion of the CR data from the
    fitted samples results in to significant narrowing of the
    experimentally available energy domain due to decreasing of the
    maximal energy boundary from $\sqrt{s}=95^{+5}_{-8}$ TeV
    \cite{TA-PRD-92-032007-2015} down to the $\sqrt{s}=13$ TeV at the
    LHC, and noticeable deterioration of approximation quality in the
    most cases. Therefore, the full data bases from Table \ref{tab:3}
    are considered below as well as corresponding fitting results from
    Table \ref{tab:4} unless otherwise stated.

One can note that the curves obtained within the hypothesis (iii)
at $\sqrt{s_{\scriptsize\mbox{min}}}=1$ TeV show some peculiar
behavior close to the lower boundary of the fitted range (Figs.
\ref{fig:1} and \ref{fig:2}). This behavior is dominated by the
standard-fit procedure, namely, by the request to obtain a better
fit quality and the absence of experimental points close to
$\sqrt{s_{\scriptsize\mbox{min}}}=1$ TeV. Therefore, this peculiar
behavior is an artificial effect due to the data analysis
procedure. Furthermore, the fit at $s \geq
s_{\scriptsize\mbox{min}}$, obviously, does not take into account
data points at lower $s$, consequently, the behavior under
consideration does not
        contradict any physical results and it has no physical meaning.
        The hypothesis (ii) is less sensitive to the absence / presence of
        the data points closer to $\sqrt{s} \approx 1$ TeV than hypothesis
        (iii). Accordingly, the curves obtained within the
        hypothesis (ii) at $\sqrt{s_{\scriptsize\mbox{min}}}=1$ TeV shows a
        smooth behavior without any features at low energy edge (thick
        dashed lines in Figs. \ref{fig:1} and \ref{fig:2}) whereas the
        curves obtained using hypothesis (iii) at
        $\sqrt{s_{\scriptsize\mbox{min}}}=1$ TeV shows the peculiar behavior at
        low energy edge (thin dashed lines in Figs. \ref{fig:1} and
        \ref{fig:2}) due to the aforementioned reasons.

    In summary, for this subsection, simultaneous fit results obtained for the full data bases DB20, DB20$_{1}$+ and DB20$_{2}$+ as well as for the corresponding ensembles of accelerator data
        DB$_{\scriptsize{\mbox{ac}}}$20,
        DB$_{\scriptsize{\mbox{ac}}}$20$_{1}$+ and
        DB$_{\scriptsize{\mbox{ac}}}$20$_{2}$+ indicates, in general, the
        robustness of the fit results for each hypothesis under
        consideration for fixed $s_{\scriptsize\mbox{min}}$, and on the
        value of the lower boundary of fitted range for
        $\sqrt{s_{\scriptsize\mbox{min}}} \geq 0.06$ TeV within certain
        hypotheses (Tables \ref{tab:4} and \ref{tab:5}). As a consequence, the corresponding curves are closer to each other in Figs.
        \ref{fig:1} and \ref{fig:2}.

    \subsection{Simultaneous Fits for the Set $\mathcal{G}_{2}$}\label{subsec:exp-1-2}

\begin{figure}
\centering
\includegraphics[width=17.0cm,height=17.0cm]{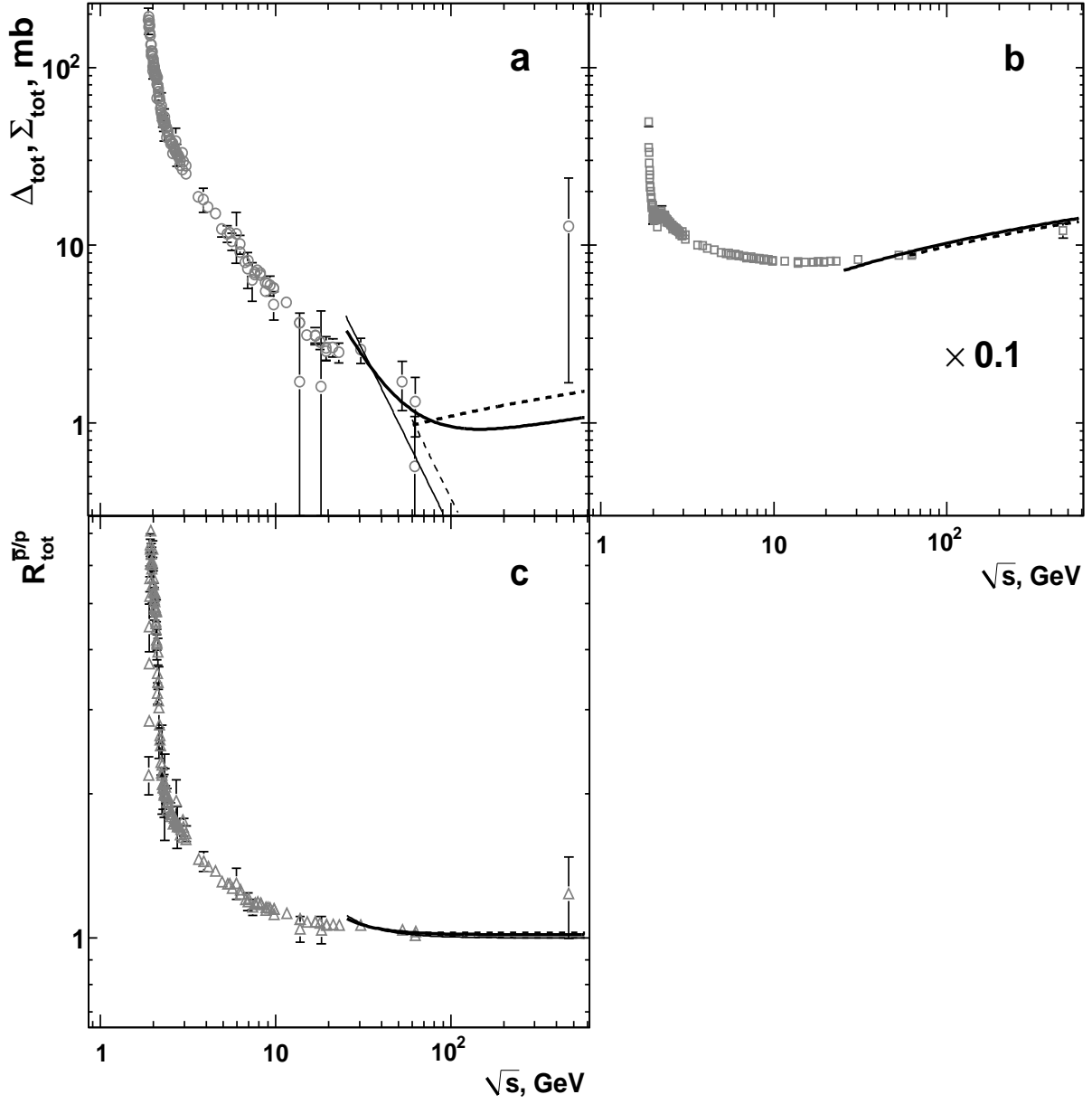}
\caption{The energy dependence of the measurements for the terms
of $\mathcal{G}_{2}$ and results of simultaneous fits of
all these terms. Points are calculated with help of the
experimental results from the DB20. The point values and fit
curves are multiplied on 0.1 for
$\Sigma_{\scriptsize{\mbox{tot}}}$ (b). The solid lines correspond
to the fit at $\sqrt{s_{\scriptsize\mbox{min}}}=0.03$ TeV, the
dashed lines are at $\sqrt{s_{\scriptsize\mbox{min}}}=0.06$ TeV.
The thick lines show the fit curves for the hypothesis (ii) and
two other lines correspond to the hypothesis (iii).} \label{fig:3}
\end{figure}

    Fig. \ref{fig:3} shows the energy dependence for
    $\mathcal{G}_{2}$, namely $\Delta_{\scriptsize{\mbox{tot}}}$ (a),
    $\Sigma_{\scriptsize{\mbox{tot}}}$ (b) and
    $R_{\scriptsize{\mbox{tot}}}^{\bar{p}/p}$ (c), with the values of
    points multiplied on 0.1 in Fig. \ref{fig:3}b for
    $\Sigma_{\scriptsize{\mbox{tot}}}(s)$ in order to use the one
    scale for the $Y$--axis for the terms $\mathcal{G}_{2}^{1}$ and
    $\mathcal{G}_{2}^{2}$. It can be seen from the plots for terms of
    $\mathcal{G}_{2}$ that values of any term $\mathcal{G}_{2}^{i}$,
    $i=1-3$ decrease at $\sqrt{s} \lesssim 10$ GeV, especially fast at
    $\sqrt{s} < 3$ GeV. For higher energies the slowdown is observed
    in the decrease for $\Delta_{\scriptsize{\mbox{tot}}}$ (Fig.
    \ref{fig:3}a) and $R_{\scriptsize{\mbox{tot}}}^{\bar{p}/p}$ (Fig.
    \ref{fig:3}c), especially noticeable in the last case; a broad
    minimum occurs for $\Sigma_{\scriptsize{\mbox{tot}}}$ (Fig.
    \ref{fig:3}b), followed by a moderate increase at $\sqrt{s}
    \gtrsim 20$ GeV.

    One can note
    that the values of $\Delta_{\scriptsize{\mbox{tot}}}$ and
    $R_{\scriptsize{\mbox{tot}}}^{\bar{p}/p}$ at highest available
    energy are larger than the previous measurements, especially in
    the first case. However, since this is the only point for each
    parameter $\Delta_{\scriptsize{\mbox{tot}}}$ and
    $R_{\scriptsize{\mbox{tot}}}^{\bar{p}/p}$, this increasing can be
    only considered as an indication on a possible transition to
    growth at $\sqrt{s} > 62$ GeV, i.e. at energies larger than the
    maximum energy of ISR, in the corresponding energy dependence.
    Except for three points with large errors
    $\Delta_{\scriptsize{\mbox{tot}}} \geq 1-2$ mb;
    $R_{\scriptsize{\mbox{tot}}}^{\bar{p}/p} > 1$ although it
    approaches to the asymptotic value (\ref{eq:asymp_1b}) from above
    with an accuracy level better than 2\% at ISR energies. However,
    indications on the change in the behavior of the
    $\Delta_{\scriptsize{\mbox{tot}}}(s)$ and
    $R_{\scriptsize{\mbox{tot}}}^{\bar{p}/p}(s)$ at higher energies,
    the absence of an exact reaching of the asymptotic levels and
    continuation of this trend at any higher energies, as well as
    other numerous studies, make it possible to exclude the reach of
    the asymptotic regime at ISR energies.

    The energy dependence of terms of $\mathcal{G}_{2}$ is
    approximated at $\sqrt{s_{\scriptsize\mbox{min}}}=3, 5, 10, 15,
    20, 25, 30, 40, 50, 60$ and 100 GeV by the corresponding formulas
    from Table \ref{tab:1} within hypothesis (ii) and (iii). It should
    be noted that the fitted samples and, as consequence, numerical
    values of the fit parameters are identical for the pairs (25, 30)
    GeV and (40, 50) GeV of the values of
    $\sqrt{s_{\scriptsize\mbox{min}}}$. Due to detailed analysis, one
    obtains that analytic formulas from Table \ref{tab:1} describe
    quite poor the energy dependence of terms of $\mathcal{G}_{2}$ at
    most of values of $\sqrt{s_{\scriptsize\mbox{min}}}$ for
    hypotheses (ii) amd (iii). The qualitative analysis indicates that
    large values of $\chi^{2}/\mbox{n.d.f.}$ are mostly dominated by
    the large discrepancy between fit curves and data points for
    $\Sigma_{\scriptsize{\mbox{tot}}}(s)$. For both hypotheses, the
    value of $\chi^{2}/\mbox{n.d.f.}$ decreases fast at growth of the
    low boundary of fitted range and the qualitative agreement is
    achieved for fitted curves and data points at
    $\sqrt{s_{\scriptsize\mbox{min}}} \geq 25$ GeV. Reasonable fit
    qualities are only observed for $\sqrt{s_{\scriptsize\mbox{min}}}
    \geq 40$ GeV. As well as for $\mathcal{G}_{1}$ such relation
    between fits and data can be considered as expected because the
    formulas from Table \ref{tab:1} are asymptotic, especially for the
    hypothesis (iii). On the other hand, experimental data for
    $\mathcal{G}_{2}$ are only available for $\sqrt{s} < 0.5$ TeV.
    This energy range is significantly narrower even than that for
    $\mathcal{G}_{1}$ and such collision energies are far from any
    estimation for the onset of the asymptotic region. Thus the
    discussion below is focused on the results for
    $\sqrt{s_{\scriptsize\mbox{min}}} \geq 0.03$ TeV with taking into
    account the identity of data samples for
    $\sqrt{s_{\scriptsize\mbox{min}}}=25$ and 30 GeV noted above.

    Fig. \ref{fig:3} shows the results of the simultaneous fits of
    corresponding data samples by using (\ref{eq:2.5}) and
    (\ref{eq:asymp_1b}) as solid (dashed) lines for
    $\sqrt{s_{\scriptsize\mbox{min}}}=0.03$ (0.06) TeV. The thick
    lines show the fit curves for the hypothesis (ii) and the two
    other lines correspond to the hypothesis (iii). The curves are
    also multiplied on 0.1 in Fig. \ref{fig:3}b in order to correspond
    to the scaled data points for
    $\Sigma_{\scriptsize{\mbox{tot}}}(s)$. The numerical values of fit
    parameters are shown in Table \ref{tab:6} for some
    $\sqrt{s_{\scriptsize\mbox{min}}} \geq 0.03$ TeV. For the
    hypothesis (ii) the fit is impossible at highest
    $\sqrt{s_{\scriptsize\mbox{min}}}=0.10$ TeV considered in this
    section due to lack of the required number of data points.

    As seen in Table \ref{tab:6}, the values of $c_{p}$ and
    $c_{\bar{p}}$ agree with each other within 1.25 standard deviation
    or better at $\sqrt{s_{\scriptsize\mbox{min}}} \geq 0.03$ TeV,
    i.e. for the fits described data points, at least, qualitatively.
    The values of $c_{p}$ and $c_{\bar{p}}$ decrease continuously with
    growth of $\sqrt{s_{\scriptsize\mbox{min}}}$, with exception of
    $c_{\bar{p}}$ at highest available
    $\sqrt{s_{\scriptsize\mbox{min}}}=0.06$ TeV, that coincides with
    the value of the parameter at
    $\sqrt{s_{\scriptsize\mbox{min}}}=0.05$ TeV. The similar situation
    is observed in Table \ref{tab:6} for $c$ parameter in the case of
    the hypothesis (iii).

    It is not possible to identify any trend in the behavior of
    $a_{1}$ depending on $s_{\scriptsize\mbox{min}}$ due to the small
    number of obtained values of the free parameter for hypothesis
    (ii). For other hypothesis studied here, $a_{1}$ is almost
    constant within uncertainties with exception the value at highest
    available $\sqrt{s_{\scriptsize\mbox{min}}}=0.10$ TeV. Detailed
    analysis shows that the fit results are not stable enough at
    highest available $s_{\scriptsize\mbox{min}}$ for both hypotheses.
    The corresponding approximated curves for
    $\Delta_{\scriptsize{\mbox{tot}}}(s)$ and
    $R_{\scriptsize{\mbox{tot}}}^{\bar{p}/p}(s)$ can show very sharp
    behavior with clear contradictions to both the data points and the
    general trends in the energy dependence of
    $\Delta_{\scriptsize{\mbox{tot}}}$ and
    $R_{\scriptsize{\mbox{tot}}}^{\bar{p}/p}$. This situation is
    similar to that observed for the set $\mathcal{G}_{1}$ and
    discussed above in subsec. \ref{subsec:exp-1-1}. Therefore, by
    analogy with the study of $\mathcal{G}_{1}$, Table \ref{tab:6}
    shows the values of fitted parameters obtained accounting for the
    additional request of smooth behavior of curves and their
    qualitative agreement with nearest data points at smaller
    collision energies for $\sqrt{s_{\scriptsize\mbox{min}}}=0.06$ /
    0.10 TeV for hypothesis (ii) / (iii).

    Fig. \ref{fig:3}a shows the curve for
    $\Delta_{\scriptsize{\mbox{tot}}}$ obtained from the simultaneous
    fit at $\sqrt{s_{\scriptsize\mbox{min}}}=0.03$ TeV within
    hypothesis (ii) corresponds at qualitative level to the main
    features of the energy dependence of data points, namely, the
    decrease of $\Delta_{\scriptsize{\mbox{tot}}}(s)$ at $\sqrt{s}
    \lesssim 60$ GeV and the possible increase of this parameter at
    higher collision energies, despite of large
    $\chi^{2}/\mbox{n.d.f.}$ On the other hand the formulas in Table
    \ref{tab:1} for $\Delta_{\scriptsize{\mbox{tot}}}(s)$ and
    $\Sigma_{\scriptsize{\mbox{tot}}}(s)$ within hypothesis (iii) show
    only smooth decrease ($ \propto \varepsilon^{-1}$) or increase
    ($\propto \ln \varepsilon$) respectively without any dependence on
    fitted energy range and, as consequence, it is difficult to
    describe the change of behavior of the $s$--dependence of terms of
    the set $\mathcal{G}_{2}$ within hypothesis (iii) as seen most
    clear in Fig. \ref{fig:3}a for
    $\Delta_{\scriptsize{\mbox{tot}}}(s)$. The fitted curves for
    $\Sigma_{\scriptsize{\mbox{tot}}}$ (Fig. \ref{fig:3}b) and
    $R_{\scriptsize{\mbox{tot}}}^{\bar{p}/p}$ (Fig. \ref{fig:3}c) are
    (very) close to each other at
    $\sqrt{s_{\scriptsize\mbox{min}}}=0.03$ and 0.06 TeV within
    certain hypothesis and for various hypotheses at fixed
    $s_{\scriptsize\mbox{min}}$.

    Summarizing, the detailed analysis of the energy dependence of terms of the set  $\mathcal{G}_{2}$ excludes the possibility of an asymptotic regime at ISR energies, and that agrees with numerous studies. The analytic functions
        deduced within the hypotheses (ii) and (iii) for the energy dependence
        of terms $\{\mathcal{G}_{2}^{i}\}_{i=1}^{3}$ describe the
        experimental data at qualitative level for
        $\sqrt{s_{\scriptsize\mbox{min}}} \geq 40$ GeV only. The curves for $\Delta_{\scriptsize{\mbox{tot}}}(s)$
        obtained with the help of the simultaneous fits for $\mathcal{G}_{2}$
        are sensitive for both hypotheses and the value of
        $s_{\scriptsize\mbox{min}}$ (Fig. \ref{fig:3}a). In contrast,
        simultaneous fit results lead to the curves for
        $\Sigma_{\scriptsize{\mbox{tot}}}$ (Fig. \ref{fig:3}b) and
        $R_{\scriptsize{\mbox{tot}}}^{\bar{p}/p}$ (Fig. \ref{fig:3}c)
        which are almost independent on the hypothesis type for a fixed
        $s_{\scriptsize\mbox{min}}$ and the value of
        $s_{\scriptsize\mbox{min}}$ for certain hypotheses.

    \begin{table*}
        \caption{\label{tab:6}Parameters for simultaneous fitting of
            $\mathcal{G}_{2}(s)$ within various hypotheses.}
        \begin{center}
                \begin{tabular}{lcccccccccc}
                    \hline \multicolumn{1}{l}{$\sqrt{s_{\scriptsize{\mbox{min}}}}$,} &
                    \multicolumn{4}{c}{hypothesis (ii)} & \multicolumn{3}{c}{} &
                    \multicolumn{3}{c}{hypothesis (iii)} \rule{0pt}{10pt}\\
                    \cline{2-11}
                    TeV & $c_{p}$, mbarn & $c_{\bar{p}}$, mbarn & $a_{2}$, mbarn & $\chi^{2}/\mbox{n.d.f.}$ & & & & $c$, mbarn & $a_{1}$, mbarn & $\chi^{2}/\mbox{n.d.f.}$ \rule{0pt}{10pt}\\
                    \hline
                    0.03 & $8.72 \pm 0.03$ & $8.78 \pm 0.03$ & $-900 \pm 200$ & $398/12$ & & & & $8.750 \pm 0.024$ & $(-1.27 \pm 0.12) \times 10^{3}$ & $404/13$\rule{0pt}{10pt}\\
                    0.05 & $8.49 \pm 0.06$ & $8.46 \pm 0.06$ & $(-2.7 \pm 0.5) \times 10^{3}$ & $31.3/9$ & & & & $8.475 \pm 0.028$ & $(-2.1 \pm 0.4) \times 10^{3}$ & $31.4/10$ \\
                    0.06 & $8.34 \pm 0.04$ & $8.44 \pm 0.04$ & $-0.15 \pm 0.07$               & $7.64/6$ & & & & $8.39 \pm 0.03$ & $(-1.9 \pm 0.5) \times 10^{3}$ & $8.17/7$ \\
                    0.10 & & -- & & & & & & $7.7 \pm 0.7$ & $(-4.5 \pm 0.6) \times 10^{4}$ & $2.17/1$ \\
                    \hline
                \end{tabular}
        \end{center}
    \end{table*}

    \subsection{Consideration for the Joined Ensemble $\mathcal{G}$}\label{subsec:exp-1-3}

    The simultaneous fit of the terms of $\mathcal{G}$ is not possible
    because, as emphasized in Sec. \ref{sec:expDB}, the databases for
    the terms of $\mathcal{G}_{2}$ are calculated from the
    experimental values of some parameters of $\mathcal{G}_{1}$,
    namely, from the measurements\footnote{One can note this situation
        principally differs from the situation for databases for
        $\sigma_{\footnotesize\mbox{tot}}$, elastic
        ($\sigma_{\footnotesize\mbox{el}}$) and inelastic
        ($\sigma_{\footnotesize\mbox{inel}}$) cross sections which are
        related by the optical theorem. There are number of experiments
        measured $\sigma_{\footnotesize\mbox{inel}}$ directly, for
        instance, \cite{PRL-109-062002-2012}. Therefore the databases for
        the set
        $\{\sigma_{\footnotesize\mbox{tot}},\sigma_{\footnotesize\mbox{el}},\sigma_{\footnotesize\mbox{inel}}\}$
        can be considered as, at least, particularly independent.} of
    $\sigma_{\footnotesize\mbox{tot}}^{xp}$, $x=p, \bar{p}$. On the
    other hand, all terms of the joined ensemble $\mathcal{G}$ are
    defined by the one set of the free parameters within certain
    hypothesis (ii) or (iii) suggested within the present work.
    Therefore, the energy dependence for the terms of one set from
    $\mathcal{G}_{1}$, $\mathcal{G}_{2}$ can be calculated with help
    of the values of free parameters obtained from the simultaneous
    fit of the terms of the another set called below as "adjoint set",
    i.e. the curves for smooth energy dependence of the terms of
    $\mathcal{G}_{1}$ at fixed $s_{\scriptsize\mbox{min}}$ can be
    calculated with the free parameter values obtained for
    $\mathcal{G}_{2}$ by the simultaneous fit at the same
    $s_{\scriptsize\mbox{min}}$, and vice versa.

\begin{figure}
\centering
\includegraphics[width=17.0cm,height=17.0cm]{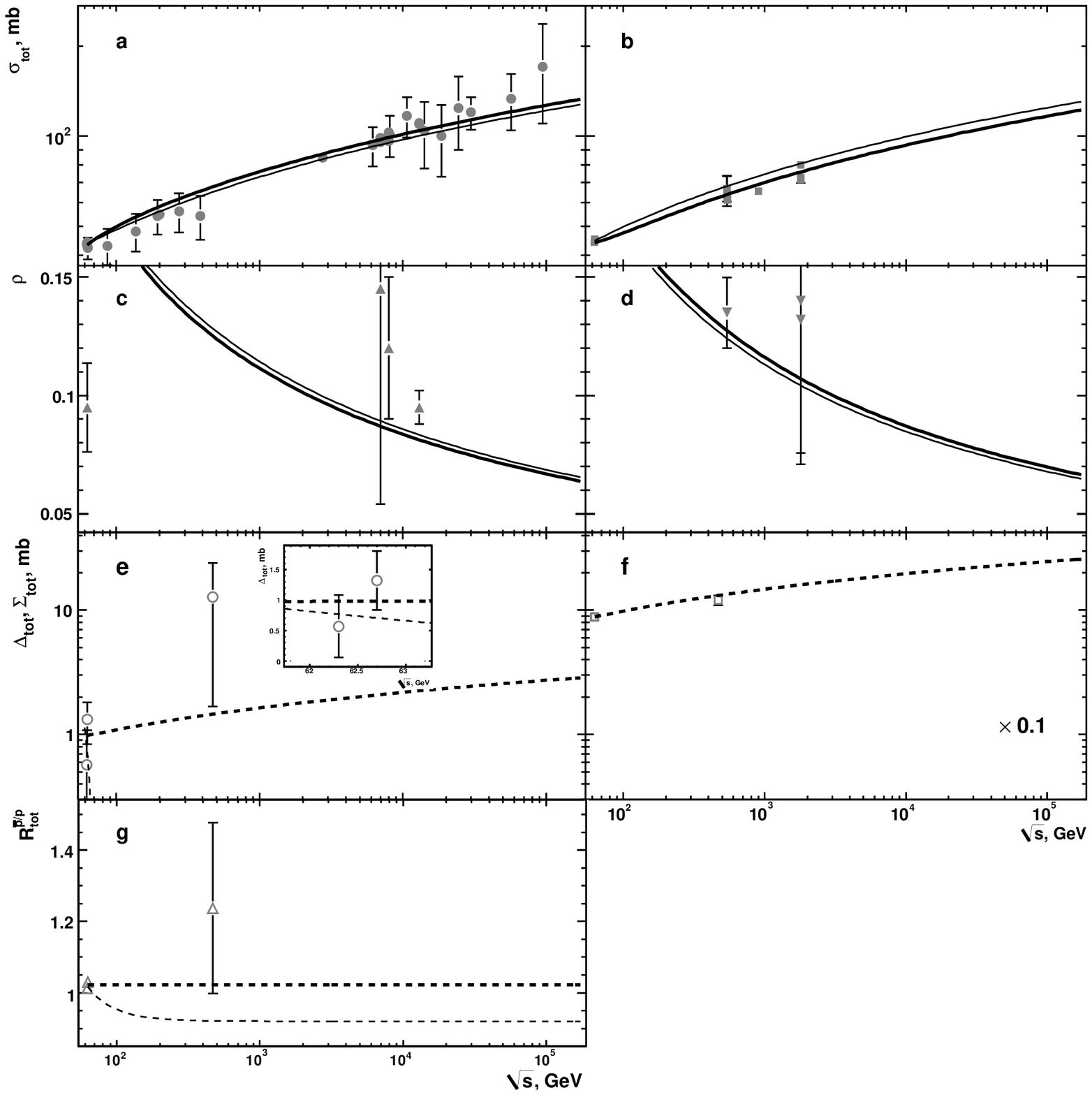}
\caption{The energy dependence for the terms of the joined set
$\mathcal{G}$ and curves obtained with help of the results shown
in Table \ref{tab:4} for the database DB20$_{1}$+ and in Table
\ref{tab:6} within hypothesis (ii). The experimental results and
curves are shown at $\sqrt{s} \geq 0.06$ TeV. The inner panel for
$\Delta_{\scriptsize{\mbox{tot}}}$ (e) shows the narrow energy
range close to the $\sqrt{s}=62.5$ GeV. The point values and
curves are multiplied on 0.1 for
$\Sigma_{\scriptsize{\mbox{tot}}}$ (f). The solid lines correspond
to the curves for the terms of $\mathcal{G}_{1}$, dashed lines are
for $\mathcal{G}_{2}$. The thick curves are from the simultaneous
fit results for the corresponding $\mathcal{G}_{i}$, $i=1$, 2; the
thin lines are calculated with help of the values of free
parameters obtained from the simultaneous fit of the terms of the
adjoint set, i.e. the curves for the terms of $\mathcal{G}_{1}$
are calculated with the free parameter values obtained for
$\mathcal{G}_{2}$ by the simultaneous fit at
$\sqrt{s_{\scriptsize\mbox{min}}}=0.06$ TeV and vice versa.}
\label{fig:4}
\end{figure}

    Fig. \ref{fig:4} shows the energy dependence for the terms of
    $\mathcal{G}$ derived within hypothesis (ii) at
    $\sqrt{s_{\scriptsize{\mbox{min}}}}=0.06$ TeV. Database
    DB20$_{1}$+ is used for the terms of $\mathcal{G}_{1}$. The thick
    curves are from the simultaneous fits for the terms of
    $\mathcal{G}_{i}$, $i=1$, 2 whereas the thin lines correspond to
    the results of the calculations for some term from
    $\mathcal{G}_{i}$, $i=1$, 2 with help of the values of free
    parameters obtained for the adjoined set $\mathcal{G}_{j}$, $j \ne
    i$ by simultaneous fit and shown in Tables \ref{tab:4} and
    \ref{tab:6} for $\sqrt{s_{\scriptsize{\mbox{min}}}}=0.06$ TeV.
    Lines corresponded to the fit results and calculations are close
    to each other for $\mathcal{G}_{1}$ (Figs. \ref{fig:4}a -- d), but
    the situation is different for most terms of $\mathcal{G}_{2}$.
    There is dramatic discrepancy between fitted and calculated curves
    for $\Delta_{\scriptsize{\mbox{tot}}}(s)$ in Fig. \ref{fig:4}e
    whereas the parameter $\Sigma_{\scriptsize{\mbox{tot}}}$ is not
    sensitive to the technique of the creation of smooth curve at all
    (Fig. \ref{fig:4}f); two curves show the similar behavior in
    functional sense at $\sqrt{s} \geq 0.2$ TeV, but there is
    quantitative difference between them for
    $R_{\scriptsize{\mbox{tot}}}^{\bar{p}/p}(s)$ in Fig. \ref{fig:4}g.

\begin{figure}
\centering
\includegraphics[width=17.0cm,height=17.0cm]{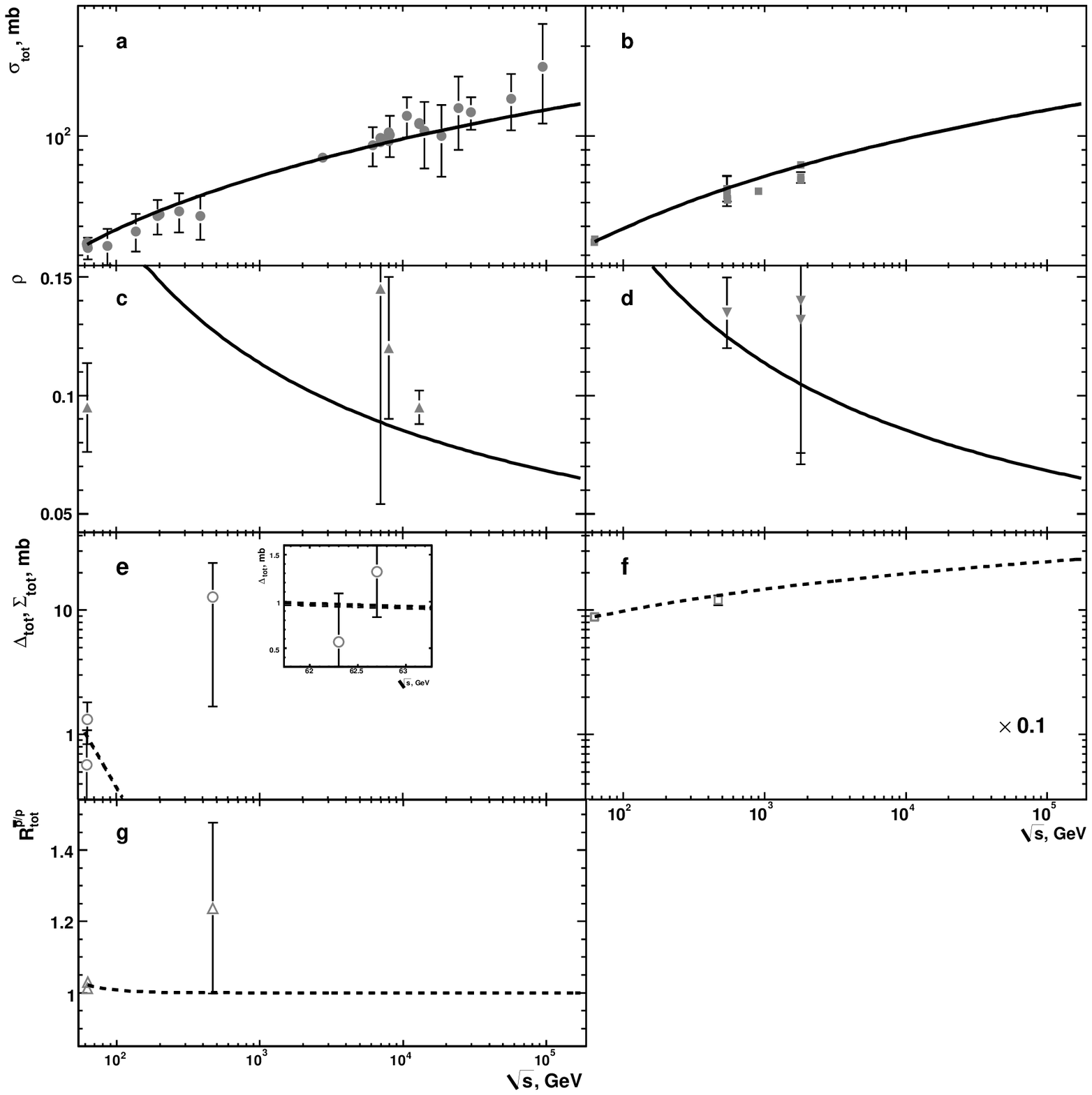}
\caption{The energy dependence for the terms of the joined set
$\mathcal{G}$ and curves obtained with help of the results shown
in Table \ref{tab:4} for the database DB20$_{1}$+ and in Table
\ref{tab:6} within hypothesis (iii). The experimental results and
curves are shown at $\sqrt{s} \geq 0.06$ TeV. The inner panel for
$\Delta_{\scriptsize{\mbox{tot}}}$ (e) shows the narrow energy
range close to the $\sqrt{s}=62.5$ GeV. The point values and
curves are multiplied on 0.1 for
$\Sigma_{\scriptsize{\mbox{tot}}}$ (f). The solid lines correspond
to the curves for the terms of $\mathcal{G}_{1}$, dashed lines are
for $\mathcal{G}_{2}$. The thick curves are from the simultaneous
fit results for the corresponding $\mathcal{G}_{i}$, $i=1$, 2; the
thin lines are calculated with help of the values of free
parameters obtained from the simultaneous fit of the terms of the
adjoint set, i.e. the curves for the terms of $\mathcal{G}_{1}$
are calculated with the free parameter values obtained for
$\mathcal{G}_{2}$ by the simultaneous fit at
$\sqrt{s_{\scriptsize\mbox{min}}}=0.06$ TeV and vice versa.}
\label{fig:5}
\end{figure}

    These features for the terms of $\mathcal{G}_{2}$ are driven by
    the unstable behavior of the fit results for $\mathcal{G}_{2}$ at
    $\sqrt{s_{\scriptsize{\mbox{min}}}}=0.06$ TeV within hypothesis
    (ii) as well as the special request added for the simultaneous fit
    for $\mathcal{G}_{2}$ described in the subsec.
    \ref{subsec:exp-1-2}. This suggestion is confirmed in Fig.
    \ref{fig:5}, that shows the energy dependence for the terms of
    $\mathcal{G}$ derived within hypothesis (iii) at
    $\sqrt{s_{\scriptsize{\mbox{min}}}}=0.06$ TeV. In this case the
    fit results for $\mathcal{G}_{2}$ are stable and additional
    request are not used for the fit procedure at all. Fig.
    \ref{fig:5} shows the full identity of the fitted and calculated
    curves for each term of the joined ensemble $\mathcal{G}$ due to
    agreement of the values of free parameters within errors in Table
    \ref{tab:4} and \ref{tab:6} for hypothesis (iii) at
    $\sqrt{s_{\scriptsize{\mbox{min}}}}=0.06$ TeV.

    It should be noted from Figs. \ref{fig:4} and \ref{fig:5} that, in
    general, one can obtain the smooth energy dependence for the terms
    of $\mathcal{G}_{1}$ in multi-TeV range with help of the fit
    results for $\mathcal{G}_{2}$ at much smaller collision energies.
    Moreover, the calculated curves for $\mathcal{G}_{1}$ agree with
    both the experimental points and the corresponding fitted curves,
    at least, reasonably. Thus, the trends in the $s$--dependence of
    the terms of $\mathcal{G}_{2}$ observed at $\sqrt{s} \lesssim 0.5$
    TeV and driven by the behavior of the
    $\sigma_{\footnotesize\mbox{tot}}^{xp}(s)$, $x=p, \bar{p}$ allow
    us to obtain the correct energy dependence for the terms of
    $\mathcal{G}_{1}$ in much wider energy domain up to the highest
    $s$ available in experiments.

    Summarizing this subsection: the energy dependence for each term of the joined set $\mathcal{G}$ can be calculated using the free parameter values obtained for any separate set $\mathcal{G}_{i}$, $i=1$, 2 by the
        simultaneous fit within the validity of the additional request of
        stability for the results of that simultaneous fit for
        $\mathcal{G}_{i}$, $i=1$, 2.

    \section{Discussion and Projections for Global Scattering Parameters}\label{sec:5}

    First of all, the results shown in Table \ref{tab:4} allow the
    qualitative estimation of the onset of asymptotic energy domain
    $s_{a}$. As seen along the text, $c_{p}$, $c_{\bar{p}}$ agree with
    $c$ within errors as well as $a_{2}$ with $a_{1}$ for simultaneous
    fits within hypotheses (ii) and (iii) at
    $\sqrt{s_{\scriptsize\mbox{min}}} \geq 5$ TeV. Thus, the fitting
    results can be considered as an evidence for possible transition
    from the hypothesis (ii) to (iii) at $\sqrt{s}\gtrsim 5$ TeV.
    Undoubtedly, this transition imposes a small energy-dependence on
    the parameters involved in the fitting procedures, not considered
    here.

    Furthermore the term $\propto \varepsilon^{-1}$ should be
    negligibly small in parameterization for
    $\sigma_{\scriptsize\mbox{tot}}^{xp}(s)$, $x=p, \bar{p}$ (Table
    \ref{tab:2}) in order to get the validation of the Pomeranchuk
    theorem in both the classical formulation (\ref{eq:asymp_1}) and the generalized one (\ref{eq:asymp_1b}).

    Considering the condition $a/\varepsilon \lesssim \delta$ for the
    energy range $s \geq s_{a}$ as, at least, one of the possible
    signatures of the onset of the asymptotic energy region, one can
    assume that the model under consideration makes it possible to
    estimate $s_{a}$ in order to magnitude. Here $a \equiv
    |a_{1}|=|a_{2}|$ within uncertainties for certain data base, DB20
    or DB20$_{1}$+, at fixed value of
    $\sqrt{s_{\scriptsize\mbox{min}}} \geq 5$ TeV and $\delta \ll 1$
    mbarn is the empirical boundary. Taking into account the range $a
    \sim (6.5 \times 10^{7} - 1.7 \times 10^{9})$ mbarn, which
    represents a variation of two orders in magnitude, and the choice
    $\delta=0.1$ mbarn, a small contribution coming from the
    non-logarithmic term, one can estimate $\sqrt{s_{a}} \sim 25.5 -
    130$ TeV. The lower value of the range for $\sqrt{s_{a}}$ is close
    to the nominal energy of the high--energy LHC (HE--LHC) mode
    \cite{FCC-CDR-V4-2018} while the upper one agrees quite reasonably
    with estimation deduced within approaches mentioned in Sec.
    \ref{sec:model} and \ref{sec:expDB}
    \cite{IJMPA-33-1850077-2018,PAN-81-508-2018}. Hopefully, this
    upper value can be achieved within newest option of the Future
    Circular Collider (FCC) project \cite{FCC-CDR-V3-2018} with proton
    beam energy 75 TeV.

    Of course, the above estimates for $\sqrt{s_{a}}$ are rather
    rough, taking into account the wide range for the absolute values
    of $a_{1,2}$ obtained with help of simultaneous fits in multi-TeV
    region $\sqrt{s_{\scriptsize\mbox{min}}} \geq 5$ TeV (Table
    \ref{tab:4}). Moreover, as stressed above, only experimental data
    for $pp$ are approximated at $\sqrt{s_{\scriptsize\mbox{min}}}
    \geq 5$ TeV. All of these allows only the preliminary statements
    at a qualitative level to be made regarding of the onset of the
    asymptotic energy domain.

    The analytic functions deduced within various hypotheses (Table
    \ref{tab:1}, \ref{tab:2}) and numerical fit results (Table
    \ref{tab:4}, \ref{tab:6}) allow the phenomenological projections
    for the terms of the wide set $\mathcal{G}$ of global scattering
    parameters and its derivative quantities, in particular, for
    energies $\mathcal{O}$(100 TeV) and higher. The main features for
    such predictions are described in detail elsewhere
    \cite{IJMPA-32-1750175-2017,PAN-82-134-2019}. Simultaneous fit results obtained for database DB20$_{1}$+ (Table \ref{tab:4}) allow predictions for the joined set $\mathcal{G}$ since
    DB20$_{1}$+ is one of the most complete databases considered here
    and, on the other hand, this database satisfies the general
    requirements used in the formation of databases. To obtain
    estimates at (ultra--)high energies, it seems reasonable to use the
    fitted results for $s_{\scriptsize{\mbox{min}}} \geq 1$ TeV$^{2}$.
    Therefore predictions for the joined set $\mathcal{G}$ are
    calculated and analyzed below, for
    $\sqrt{s_{\scriptsize{\mbox{min}}}}=1$, 5, and 10 TeV. Calculations
    are performed for both collisions considered in the paper
    ($pp, \bar{p}p$) for $\sqrt{s} \geq 14$ TeV.

    Within hypothesis (ii), the values of both the
    $\sigma_{\footnotesize\mbox{tot}}^{xp}$ and $\rho^{xp}$ do not
    depend on the type of collision for a given
    $s_{\scriptsize{\mbox{min}}}$, nor
    $s_{\scriptsize{\mbox{min}}}$ indicated above for a fixed type of
    interaction within errors. The $\rho^{xp}$ reaches the asymptotic
    level (\ref{eq:rho_3}) within uncertainty at smallest $\sqrt{s} =
    14$ TeV under consideration at any $s_{\scriptsize{\mbox{min}}}$.
    Relative uncertainties for the estimations of
    $\sigma_{\footnotesize\mbox{tot}}^{xp}$ ($\delta_{\sigma}^{xp}$)
    are almost constant for all $s_{\scriptsize{\mbox{min}}}$, except
    for the highest $\sqrt{s_{\scriptsize{\mbox{min}}}}=10$ TeV, for
    which, after some decrease (increase), a fairly rapid transition
    to a constant level is observed at $\sqrt{s} \approx 25$ (50) TeV
    for $pp$ ($\bar{p}p$) collisions. The accuracy of estimates for
    $pp$ is noticeably better than that for $\bar{p}p$, deteriorating for both collision types with increasing of
    $s_{\scriptsize{\mbox{min}}}$, especially for transition from
    lower values of $s_{\scriptsize{\mbox{min}}}$ to the highest
    $\sqrt{s_{\scriptsize{\mbox{min}}}}=10$ TeV under consideration.
    In the case of $pp$ ($\bar{p}p$), the values of
    $\delta_{\sigma}^{xp}$ are at the level of 0.09 (0.22) at
    $\sqrt{s_{\scriptsize{\mbox{min}}}}=1$ TeV, slightly increasing to
    0.11 (0.27) at $\sqrt{s_{\scriptsize{\mbox{min}}}}=5$ TeV. The
    decrease of $\delta_{\sigma}^{pp}$ is observed from 0.17 to 0.15
    at $\sqrt{s} \gtrsim 25$ TeV, whereas there is a smooth increase of
    $\delta_{\sigma}^{\bar{p}p}$ from 0.36 to a constant level of 0.38
    at $\sqrt{s} \gtrsim 50$ TeV at
    $\sqrt{s_{\scriptsize{\mbox{min}}}}=10$ TeV. In the case of the
    $\rho^{xp}$, in general, a similar situation is observed for
    $\delta^{xp}_{\rho}$ in the functional sense as well as for
    $\delta^{xp}_{\sigma}$: for $pp$, the $\delta^{pp}_{\rho}$ quantity
    is approximately constant for all $s_{\scriptsize{\mbox{min}}}$
    except for the $\sqrt{s_{\scriptsize{\mbox{min}}}}=10$ TeV,
    for which the approach to the constant level is observed at
    $\sqrt{s} \gtrsim 25$ TeV after some decrease of
    $\delta^{pp}_{\rho}(s)$; for $\bar{p}p$, the
    $\delta^{\bar{p}p}_{\rho}(s)$ agrees with constant quite well at
    lowest value $\sqrt{s_{\scriptsize{\mbox{min}}}}=1$ TeV under
    discussion, and relative uncertainty $\delta^{\bar{p}p}_{\rho}$
    shows some increase for other $s_{\scriptsize{\mbox{min}}}$,
    especially noticeable at highest
    $\sqrt{s_{\scriptsize{\mbox{min}}}}=10$ TeV reaching a constant
    level at $\sqrt{s} \gtrsim 50$ TeV. The relative uncertainties in projections of the $\rho$-parameter are smaller than those for the
    total cross section for any of the collision types considered here
    at any fixed $s_{\scriptsize{\mbox{min}}}$. For $pp$, the
    $\delta^{pp}_{\rho}(s)$ quantity agrees well with a constant 0.068
    for $\sqrt{s} \geq 14$ TeV at
    $\sqrt{s_{\scriptsize{\mbox{min}}}}=1$ TeV whereas some decrease
    is observed for higher $s_{\scriptsize{\mbox{min}}}$, especially
    for highest $\sqrt{s_{\scriptsize{\mbox{min}}}}=10$ TeV. The
    $\delta^{pp}_{\rho}(s)$ decreases with growth $s$ from 0.106
    (0.151) at $\sqrt{s}=14$ TeV down to the constant level 0.105
    (0.133) at $\sqrt{s} \geq 20$ (200) TeV for
    $\sqrt{s_{\scriptsize{\mbox{min}}}}=5$ (10) TeV. The
    $\delta^{\bar{p}p}_{\rho}(s)$ is approximately equal to the
    constant level $0.078$ for $\sqrt{s} \geq 14$ TeV at
    $\sqrt{s_{\scriptsize{\mbox{min}}}}=1$ TeV and
    $\delta^{\bar{p}p}_{\rho}$ increases from 0.082 (0.112) to a
    constant of 0.085 (0.127) at $\sqrt{s} \gtrsim 30$ (150) TeV for
    $\sqrt{s_{\scriptsize{\mbox{min}}}}=5$ (10) TeV. Thus, in the case
    of the $\rho$-parameter, the inverse energy dependence of the
    relative uncertainty is observed compared to that seen for the
    $\sigma_{\footnotesize\mbox{tot}}^{xp}$ at
    $\sqrt{s_{\scriptsize{\mbox{min}}}} > 1$ TeV.

    Estimates for each term of the set $\mathcal{G}_{1}$,
    calculated within the hypothesis (iii), are characterized by
    significantly better accuracy than those in the case of hypothesis
    (ii) for each of the $s_{\scriptsize{\mbox{min}}}$ used for the
    derivation of numerical estimates.

    Within the hypothesis (iii), the values of the
    $\sigma_{\footnotesize\mbox{tot}}^{xp}$ do not depend on the type
    of collision within the errors for a given
    $s_{\scriptsize{\mbox{min}}}$ as well as for the hypothesis (ii).
    However, in contrast to hypothesis (ii), some increase of
    estimations of $\sigma_{\footnotesize\mbox{tot}}^{xp}$ at fixed
    $s$ is observed with an increase in $s_{\scriptsize{\mbox{min}}}$,
    especially in the transition from $\sqrt{s_{\scriptsize{\mbox{min}}}}
    = 1$ TeV to the highest $\sqrt{s_{\scriptsize{\mbox{min}}}} = 10$
    TeV under consideration. The ratios of the estimates obtained for
    a fixed type of collision at given $s$ and different
    $s_{\scriptsize{\mbox{min}}}$ -- $\displaystyle R^{\,xp}_{\sigma,
        2/1} \equiv
    \frac{\left.\sigma_{\footnotesize\mbox{tot}}^{xp}(s)\right|_{s_{\scriptsize{\mbox{min},2}}}}
    {\left.\sigma_{\footnotesize\mbox{tot}}^{xp}(s)\right|_{s_{\scriptsize{\mbox{min},1}}}}$
    -- reach constant levels with an increase in $s$, and this occurs,
    in general, at the different $s$ in dependence on both the
    $s_{\scriptsize{\mbox{min}}}$ and the collision type. The
    numerical values of these constant levels are $\forall\,x=p,
    \bar{p}: R^{\,xp}_{\sigma, 2/1}=1.06 \pm 0.03$, $1.16 \pm 0.07$
    and $1.09 \pm 0.07$ at $\sqrt{s} \gtrsim 60$, 50 and 40 (60, 70
    and 140) TeV in $pp$ ($\bar{p}p$) interactions for the pairs of
    lower boundary values for fitted ranges
    $(\sqrt{s_{\scriptsize{\mbox{min},1}}},\sqrt{s_{\scriptsize{\mbox{min},2}}}\,)=(1,~
    5)$, (1, 10) and (5, 10) in TeV, respectively. Thus the values of
    these constants with uncertainties do not depend either on the type
    of collision or the choice of the specific pair
    $(s_{\scriptsize{\mbox{min},1}},s_{\scriptsize{\mbox{min},2}})$.
    The quantity $\delta_{\sigma}^{xp}$ agrees quite well with
    constant at $\sqrt{s} \geq 14$ TeV, and does not depend on the type
    of interaction for all $s_{\scriptsize{\mbox{min}}}$, except for
    $\sqrt{s_{\scriptsize{\mbox{min}}}} = 10$ TeV, for which the
    onset of constant behavior is observed at $\sqrt{s} \approx 100$
    TeV after some decrease from 0.075 (0.065), at the smallest
    $\sqrt{s}=14$ TeV under consideration for $pp$ ($\bar{p}p$)
    collisions. The constant values for $\delta_{\sigma}^{xp}$ are
    approximately 0.006, 0.024 and 0.061 at
    $\sqrt{s_{\scriptsize{\mbox{min}}}} = 1$, 5, and 10 TeV,
    respectively. The quantity $\delta^{xp}_{\rho}$ does not depend on
    the type of interaction and, due to the reasons indicated above in
    Sec. \ref{sec:model}, it becomes small, namely, $<
    10^{-4}$ already at $\sqrt{s} \gtrsim 30$, 125, and 350 TeV for
    $\sqrt{s_{\scriptsize{\mbox{min}}}} = 1$, 5, and 10 TeV,
    respectively. Furthermore, the $\delta^{xp}_{\rho}$ rapidly
    decreases with the increasing $s$. Therefore, within the framework of
    the hypothesis (iii), the $\delta^{xp}_{\rho}$ can be taken equal
    to zero at $\sqrt{s} \gtrsim 30$, 125, and 350 TeV for
    $\sqrt{s_{\scriptsize{\mbox{min}}}} = 1$, 5, and 10 TeV. Accounting
    for the aforementioned specific situation with the uncertainties
    for estimates of the $\rho^{xp}$-parameter, the median values of
    these estimates are only discussed below in this paragraph. The
    median values of the $\rho$-parameter are almost independent of
    the type of collision and coincide with the asymptotic level
    (\ref{eq:rho_3}) with accuracy $10^{-3}$ already at the smallest
    of the considered $\sqrt{s}=14$ TeV for
    $\sqrt{s_{\scriptsize{\mbox{min}}}} = 1$ TeV. There is no
    dependence on the $\rho$-parameter for the type of interaction, and
    the median values of $\rho^{xp}$ coincide with $\rho^{xp}_{a}$
    (\ref{eq:rho_3}) with accuracy $10^{-3}$ at $\sqrt{s} \gtrsim 50$
    (100) TeV in the case of $\sqrt{s_{\scriptsize{\mbox{min}}}} = 5$
    (10) TeV. The situation for the ratio of estimates of $\rho$
    obtained for a fixed type of interaction at given $s$ and
    different $s_{\scriptsize{\mbox{min}}}$ -- $\displaystyle
    R^{\,xp}_{\rho, 2/1} \equiv
    \frac{\left.\rho^{xp}(s)\right|_{s_{\scriptsize{\mbox{min},2}}}}
    {\left.\rho^{xp}(s)\right|_{s_{\scriptsize{\mbox{min},1}}}}$ -- in
    general, is similar to that observed for
    $\sigma_{\footnotesize\mbox{tot}}^{xp}$. But, unlike on total
    cross sections, $\forall\,x=p, \bar{p}: R^{\,xp}_{\rho, 2/1}$
    agree with unity with high accuracy better than $7 \times 10^{-4}$
    at $\sqrt{s} \geq 100$ TeV for any pair
    $(s_{\scriptsize{\mbox{min},1}},s_{\scriptsize{\mbox{min},2}})$
    considered here.

    The detailed analysis made above for the predictions for the terms of
    the set $\mathcal{G}_{1}$ unambiguously indicates that the terms
    $\Delta_{\scriptsize{\mbox{tot}}}$ and
    $R_{\scriptsize{\mbox{tot}}}^{\bar{p}/p}$ of the set
    $\mathcal{G}_{2}$ coincide with their asymptotic values
    (\ref{eq:asymp_1}) and (\ref{eq:asymp_1b}) within the uncertainties at
    $\sqrt{s} \geq 14$ TeV for both hypotheses (ii) and (iii), at any
    $s_{\scriptsize{\mbox{min}}} \geq 1$ TeV$^{2}$. The smooth
    increase is observed for $\sigma_{\footnotesize\mbox{tot}}^{xp}$
    with the growth of $s$ for any $s_{\scriptsize{\mbox{min}}}$. Therefore,
    one can conclude that the realization of the scenario according to the Pomeranchuk theorem
    (\ref{eq:asymp_1b}) seems more preferable at asymptotic energies
    than the scenario with the formulation (\ref{eq:asymp_1}) for any
    hypotheses discussed in this paper.

    Furthermore, the aforementioned study of energy behavior of
    projections for the terms of the joined ensemble $\mathcal{G}$ shows
    that, in general, physical quantities and their uncertainties reach
    corresponding asymptotic or constant levels at energies
    $\mathcal{O}$(100 TeV). That can be considered as indirect
    evidence in favor of the range for $\sqrt{s_{a}}$, derived with
    help of the condition for $a/\varepsilon$.

    \begin{table*}
        \caption{\label{tab:7} Predictions for $pp$ based on the
            simultaneous fit of $\mathcal{G}_{1}$ for DB20$_{1}$+ at
            $\sqrt{s_{\scriptsize{\mbox{min}}}}=5$ TeV with first / second
            line in the cell for the hypothesis (ii) / (iii).}
        \begin{center}
                \begin{tabular}{cccccccccc}
                    \hline \multicolumn{1}{l}{} &
                    \multicolumn{9}{c}{$\sqrt{s}$, TeV} \rule{0pt}{10pt}\\
                    \cline{2-10} \multicolumn{1}{l}{Parameter} &
                    \multicolumn{3}{c}{(HL--)LHC, HE--LHC, LHC-ultimate} &
                    \multicolumn{6}{c}{SPPC, FCC-hh, VLHC-I, II} \rule{0pt}{10pt}\\
                    & 14 & 27 & 42 & 40 & 70.6 & 100 & 125 & 150 & 175 \\
                    \hline
                    $\sigma_{\scriptsize{\mbox{tot}}}^{pp}$, mbarn & $108 \pm 12$ & $117 \pm 13$ & $122 \pm 13$ & $122 \pm 13$ & $128 \pm 14$ & $132 \pm 14$ & $135 \pm 14$ & $137 \pm 15$ & $139 \pm 15$\rule{0pt}{10pt}\\
                    & $108.5 \pm 2.7$ & $117.3 \pm 2.8$ & $122.7 \pm 2.9$ & $122.1 \pm 2.9$ & $129 \pm 3$ & $133 \pm 3$ & $135 \pm 3$ & $138 \pm 3$ & $139 \pm 3$ \rule{0pt}{10pt}\\
                    $\rho^{pp} \times 10^{3}$ & $87 \pm 9$ & $80 \pm 8$ & $77 \pm 8$ & $77 \pm 8$ & $73 \pm 8$ & $71 \pm 7$ & $69 \pm 7$ & $68 \pm 7$ & $67 \pm 7$ \rule{0pt}{10pt}\\
                    & $83.6 \pm 0.4$ & $77.3 \pm 0.1$ & $73.89 \pm 0.03$ & $74.24 \pm 0.04$ & $70.38 \pm 0.01$ & $68.24 \pm 0.01$ & $66.93$ & $65.91$ & $65.06$ \rule{0pt}{10pt}\\
                    \hline \multicolumn{1}{l}{} & \multicolumn{1}{c}{VLHC-II} &
                    \multicolumn{6}{c}{ultra-high energy
                        cosmic rays} & \multicolumn{2}{c}{higher energies} \rule{0pt}{10pt}\\
                    & 200 & 110 & 170 & 250 & 500 & 750 & $10^{3}$ & $5 \times 10^{3}$ & $10^{4}$ \\
                    \hline
                    $\sigma_{\scriptsize{\mbox{tot}}}^{pp}$, mbarn & $140 \pm 15$ & $133 \pm 14$ & $138 \pm 15$ & $143 \pm 15$ & $151 \pm 16$ & $155 \pm 17$ & $159 \pm 17$ & $172 \pm 19$ & $185 \pm 20$ \rule{0pt}{10pt}\\
                    & $141 \pm 3$ & $134 \pm 3$ & $139 \pm 3$ & $143 \pm 3$ & $151 \pm 4$ & $156 \pm 4$ & $159 \pm 4$ & $178 \pm 4$ & $186 \pm 4$ \rule{0pt}{10pt}\\
                    $\rho^{pp} \times 10^{3}$ & $67 \pm 7$ & $70 \pm 7$ & $68 \pm 7$ & $66 \pm 7$ & $62 \pm 7$ & $60 \pm 6$ & $59 \pm 6$ & $53 \pm 6$ & $51 \pm 5$ \rule{0pt}{10pt}\\
                    & $64.35$ & $67.67$ & $65.22$ & $63.19$ & $59.85$ & $58.06$ & $56.85$ & $50.92$ & $48.73$ \rule{0pt}{10pt}\\
                    \hline
                \end{tabular}
        \end{center}
    \end{table*}

    Some numerical values for estimates of global scattering
    parameters are shown in Table \ref{tab:7} for $pp$ in the energy
    range from the nominal $\sqrt{s}$ of the LHC up to the high
    boundary of the PeV domain $\sqrt{s}=10$ PeV based on the detailed
    analysis above and the reasons discussed elsewhere
    \cite{IJMPA-32-1750175-2017,PAN-82-134-2019}. The fitted results
    for DB20$_{1}$+ at $\sqrt{s_{\scriptsize{\mbox{min}}}}=5$ TeV are
    used to calculate the estimates of terms of the set
    $\mathcal{G}_{1}$. The choice of the value
    $\sqrt{s_{\scriptsize{\mbox{min}}}}=5$ TeV for Table \ref{tab:7}
    is based on the detailed analysis in subsec. \ref{subsec:exp-1-1} as
    well as in the present section. In particular, the simultaneous fit
    for $\mathcal{G}_{1}$ at that $s_{\scriptsize{\mbox{min}}}$ is
    characterized firstly by enough number of experimental points in
    fitted sample together with statistically acceptable
    $\chi^{2}/\mbox{n.d.f.}$ in difference with smaller
    $s_{\scriptsize{\mbox{min}}}$ and, secondly, by the higher
    robustness of fit results in comparison with the
    $\sqrt{s_{\scriptsize{\mbox{min}}}}=10$ TeV. As seen from Table
    \ref{tab:7}, the predictions for
    $\sigma_{\scriptsize{\mbox{tot}}}^{pp}$ and $\rho^{pp}$ coincide
    within the uncertainties for various hypotheses under consideration.
    Direct comparison is impossible for the results in Table
    \ref{tab:7} and the projections within AQFT
    \cite{IJMPA-32-1750175-2017} because of noticeably different
    $s_{\scriptsize{\mbox{min}}}$ used for simultaneous fits and, consequently, for calculations of estimates for
    $\sigma_{\scriptsize{\mbox{tot}}}^{pp}$ and $\rho^{pp}$.
    Nevertheless, accounting for this feature, one can observe the
    following: the estimates obtained within the present work are
    characterized much better precision than those within AQFT
    \cite{IJMPA-32-1750175-2017}, especially for $\rho$-parameter; the
    estimates mostly agree within uncertainties for two approaches for
    any global scattering parameters under discussion, except for
    of $\sigma_{\scriptsize{\mbox{tot}}}^{pp}$ at ultra--high energies
    $\sqrt{s} \geq 5$ PeV. In general, this discrepancy can be expected
    since the present approach provides
    $\sigma_{\scriptsize{\mbox{tot}}}^{pp} \propto \ln\varepsilon$ at
    $s \to \infty$, i.e. the slowest increase of the total cross
    section at (ultra--)high energies, and this functional form for
    $\sigma_{\scriptsize{\mbox{tot}}}^{pp}$ should lead to the
    continuously increasing difference with results of AQFT with the
    growth of $s$.

    \section{Conclusion}\label{sec:6}

    The asymptotic behavior of the wide set $\mathcal{G}$ of
    scattering parameters is studied with the help of crossing
    property, the DDR and the optical theorem in $pp$ and
    $\bar{p}p$ collisions. It is important to point out that we
    obtained an intercept $\alpha_{\mathbb{P}}=1$, typical of a soft
    Pomeron (low momentum transfer). This is a consequence of the use
    of the first-order approximation for the DDR. In contrast,
    the introduction of higher derivative orders to describe the
    dispersion relation may introduce subleading contributions at high
    energies, which can lead to a Pomeron intercept different from 1.
    The simplest functional form for the real part of the forward
    scattering amplitude is used for the asymptotic energy domain to
    deduce the analytic expressions for
    $\sigma_{\footnotesize\mbox{tot}}^{xp}$ and $\rho^{xp}$, $x=p,
    \bar{p}$ as well as for some combinations of total cross sections,
    which are important for verification of the Pomeranchuk theorem.
    Within that form, for $\mathrm{Re}F_{xp}$, the two hypotheses --
    (ii) and (iii) -- are kept with non-zero constants for
    $\mathrm{Re}F_{xp}$ at asymptotic energies.

    Three consequent stages are considered for the most current database for
    global scattering parameters in elastic $pp$ and $\bar{p}p$
    scattering, namely, the latest PDG sample (DB20) and all available
    experimental results (DB20$_{1}$+, DB20$_{2}$+). The analytic
    parameterizations deduced considering both hypotheses provide a
    quantitative description of energy-dependence of measured global
    scattering parameters -- set $\mathcal{G}_{1}$ -- with mostly
    robust values of fit parameters for any DBs. The fit qualities are
    reasonable for energy range $\sqrt{s} \geq 0.06$ TeV, and the
    quantity is statistically acceptable at the low boundary for fit
    domain $\sqrt{s_{\scriptsize{\mbox{min}}}} \geq 0.1$ TeV for
    hypothesis (ii). The corresponding ranges are significantly
    shifted to the higher energies and they are $\sqrt{s} \geq 0.5$
    TeV and $\sqrt{s} \geq 1$ TeV for statistically reasonable and
    acceptable quantities, respectively, for hypothesis (iii) due to its
    ''extremely" asymptotic nature. Detailed analysis of accelerator
    data only shows that hypotheses (ii) and (iii) allow the
    qualitative description for these data up to the
    $\sqrt{s_{\scriptsize{\mbox{min}}}}=5$ TeV. The experimental database
    created for some interrelations between
    $\sigma_{\scriptsize{\mbox{tot}}}^{\bar{p}p}$ and
    $\sigma_{\scriptsize{\mbox{tot}}}^{pp}$ -- set $\mathcal{G}_{2}$
    -- covers the energy domain $\sqrt{s} < 0.5$ TeV only and most of the
    points are at $\sqrt{s} \lesssim 0.06$ TeV. The analytic
    expressions deduced for terms of the set $\mathcal{G}_{2}$ within
    hypotheses (ii) and (iii) only provide a qualitative description of
    the energy dependence of these terms at
    $\sqrt{s_{\scriptsize{\mbox{min}}}} \geq 0.04$ TeV. In general,
    such relation between simultaneous fits and data is expected since the formulas for the terms of the set $\mathcal{G}_{2}$, as
    well as for the set $\mathcal{G}_{1}$, are asymptotic for both
    hypotheses under study. Nevertheless, the consideration of the
    joined ensemble $\mathcal{G}$ indicates that, in general, one can
    obtain the smooth energy-dependence for the terms of
    $\mathcal{G}_{1}$ in multi-TeV range with the help of the fit results
    for $\mathcal{G}_{2}$ at much smaller collision energies, and the
    calculated curves for $\mathcal{G}_{1}$ agree with both the
    experimental data and the corresponding fitted curves, at least,
    reasonably.

    Study of empirical conditions for the functional forms of the
    analytic expressions for measured global scattering parameters
    allows the following rough estimation of collision energy for the
    onset of the asymptotic region $\sqrt{s_{a}} \sim 25.5 - 130$
    TeV.

    Based on the simultaneous fit results, the estimations are
    calculated for the total cross section and the $\rho$-parameter, considering the elastic $pp$ scattering at different $s$, up to energy frontier
    $\sqrt{s}=10$ PeV. These estimations with uncertainties are robust
    for various $\sqrt{s_{\scriptsize{\mbox{min}}}} \geq 1$ TeV and
    types of collisions. The numerical values of the estimations for
    $\sigma_{\scriptsize{\mbox{tot}}}^{pp}$ and $\rho^{pp}$ obtained
    with the help of the simultaneous fit of $\mathcal{G}_{1}$ for
    DB20$_{1}$+ at $\sqrt{s_{\scriptsize{\mbox{min}}}}=5$ TeV agree
    for hypothesis (ii) and (iii) within the errors. The estimations for
    $\sigma_{\scriptsize{\mbox{tot}}}^{pp}$ can be considered conservative due to functional form for this quantity at
    (ultra--)high energies which consequently leads to a slow increase of
    $\sigma_{\scriptsize{\mbox{tot}}}^{pp}$ as $s$ rise. Detailed
    analysis of the predictions for terms of the joined ensemble
    $\mathcal{G}$ unambiguously indicates that the realization of the
    scenario with the generalized formulation of the Pomeranchuk
    theorem seems more suitable at asymptotic energies than the
    scenario with the original formulation of this theorem, for any
    hypotheses discussed along the paper. The aforementioned study shows
    that, in general, physical quantities and their uncertainties reach
    corresponding asymptotic or constant levels at energies
    $\mathcal{O}$(100 TeV). As commented before, it can be considered an indirect indication to support the range for $\sqrt{s_{a}}$, obtained by considering the empirical conditions described in this work.

    As a last comment, it is important to stress that at very high energies, possible non-linearities on the elastic forward quantities may be taken into account by using higher-order derivative terms in the dispersion relations. This procedure may lead to corrections on the results presented here. This question is presently under study and will be published elsewhere.

    \section*{Acknowledgments}

    SDC thanks to UFSCar for the financial support. The work of V.A.O.
    was supported partly by NRNU MEPhI Program ''Priority 2030".

\clearpage

\clearpage

\clearpage

\clearpage

\clearpage


\begin{thebibliography}{0}    

\bibitem{froissart_phys_rev_123_1053_1961}M. Froissart, Phys. Rev. \textbf{123}, 1053 (1961).

\bibitem{martin_nuovo_cim_42_930_1966}A. Martin, II Nuovo Cim. A\textbf{42}, 930 (1966).

\bibitem{A.Martin.Nuovo.Cim.A44.1219.1966}A. Martin, Nuovo Cim. A\textbf{44}, 1219 (1966).

\bibitem{lukaszuk_nuovo_cimen_a52_122_1967}L. Lukaszuk and A. Martin. Nuovo Cim. A\textbf{52}, 122 (1967).

\bibitem{Martin-PR-D80-065013-2009}A. Martin, Phys. Rev. D\textbf{80}, 065013 (2009).

\bibitem{O.Nachtmann.Ann.Phys.209.436.1991}O. Nachtmann, Ann. Phys. \textbf{209}, 436 (1991).

\bibitem{pomeranchuk_jetp_7_499_1958}I. Y. Pomeranchuk. JETP \textbf{7}, 499 (1958).

\bibitem{S.D.Campos.Chin.Phys.C.2020}S. D. Campos, Chin. Phys. C\textbf{44}, 103103 (2020).

\bibitem{F.E.Low.Phys.Rev.D12.163.1975}F. E. Low, Phys. Rev. D\textbf{12}, 163 (1975)

\bibitem{S.Nussinov.Phys.Rev.Lett.34.1286.1975}S. Nussinov, Phys. Rev. Lett. \textbf{34}, 1286 (1975).

\bibitem{eden_book_2}R. J. Eden, P. V. Landshoff, D. I. Olive and J. C. Polkinghorne, \emph{The Analytic S-Matrix}. Cambridge Univ. Press (1966).

\bibitem{S.D.Campos.EPJC.47.171.2006}R. F. \'Avila, S. D. Campos, M. J. Menon and J. Montanha, Eur. Phys. J. C\textbf{47}, 171 (2006).


\bibitem{Collins-book-1968}
P. D. B. Collins and E. J. Squires, \emph{Regge poles in particle physics}. Springer tracts in modern physics, \textbf{45}. Springer--Verlag (1968).

\bibitem{NPA-744-249-2004}R. F. \'Avila and M. J. Menon, Nucl. Phys. A\textbf{744}, 249 (2004).

\bibitem{Donnachie-book-2002}
A. Donnachie, H. G. Dosch, P. V. Landshoff and O. Nachtmann,
\emph{Pomeron physics and QCD}. Cambridge Univ. Press (2002).

\bibitem{PTEP-2020-083C01-2020}
P.A. Zyla {\it et al.} (Particle Data Group), Prog. Theor. Exp.
Phys. \textbf{2020}, 083C01 (2020).

\bibitem{T.Kinoshita.J.J.Loeffel.A.Martin.Phys.Rev.Lett.10.460.1963}T. Kinoshita, J. J. Loeffel and A. Martin, Phys. Rev. Lett. \textbf{10}, 460 (1963).

\bibitem{eden_phys_rev_lett_16_39_1966} R. J. Eden, Phys. Rev. Lett. \textbf{16}, 39 (1966).

\bibitem{kinoshita_book_1966}T. Kinoshita, in: \textit{Perspectives in Modem Physics}, Ed. R. E. Marshak. p. 211. Wiley, New York (1966).

\bibitem{grunberg_phys_rev_lett_31_63_1973}G. Grunberg and Tran N. Truong, Phys. Rev. Lett. \textbf{31}, 63 (1973); Phys. Rev. D\textbf{9}, 2874 (1974).

\bibitem{D.A.Fagundes.A.Grau.G.Pancheri.O.Shekhovtsova.Y.N.Srivastava.Phys.Rev.D96.054010.2017} D. A. Fagundes, A. Grau, G. Pancheri, O. Shekhovtsova and Y. N. Srivastava, Phys. Rev. D\textbf{96}, 054010 (2017).

\bibitem{Leader-book-V2-1996}
E. Leader and E. Predazzi, \emph{An intriduction to gauge theories
and modern particle physics}. Vol. \textbf{2}. Cambridge Univ.
Press (1996).

\bibitem{IJMPA-33-1850077-2018}
V. A. Petrov and V. A. Okorokov, Int. J. Mod. Phys. A\textbf{33},
1850077 (2018).

\bibitem{PAN-81-508-2018}
V. A. Okorokov, Phys. At. Nucl. \textbf{81}, 508 (2018).

\bibitem{IJMPA-A25-5333-2010}
S. D. Campos and V. A. Okorokov, Int. J. Mod. Phys. A\textbf{25},
5333 (2010).

\bibitem{IJMPA-32-1750175-2017}
V. A. Okorokov and S. D. Campos, Int. J. Mod. Phys. A\textbf{32},
1750175 (2017).

%

%

%

\bibitem{PLB-808-135663-2020}
J. Adam {\it et al.} (STAR Collaboration), Phys. Lett.
C\textbf{808}, 135663 (2020).

\bibitem{EPJC-79-785-2019}
G. Antchev \emph{et al.} (TOTEM Collaboration), Eur. Phys. J.
C\textbf{79}, 785 (2019).

\bibitem{okorokov-arxiv-1501.01142} V. A. Okorokov, Adv. High Energy
Phys. \textbf{2015}, 914170 (2015); the database for the forward
slope is available at arXiv: 1501.01142 [hep-ph] (2015).

\bibitem{Okorokov-IJMPA-27-1250037-2012}
V. A. Okorokov, Int. J. Mod. Phys. \textbf{A27}, 1250037 (2012).

\bibitem{Okorokov-AHEP-2015-790646-2015}
V. A. Okorokov, Adv. High Energy Phys. \textbf{2015}, 790646
(2015); \textit{ibid} \textbf{2016}, 5972709 (2016); \textit{ibid}
\textbf{2017}, 5465398 (2017).

\bibitem{M.M.Block.R.N.Cahn.Phys.Lett.B149.245.1984}M. M. Block and R. N. Cahn, Phys. Lett. B\textbf{149}, 245 (1984); Rev. Mod. Phys. \textbf{57}, 563 (1985).

\bibitem{T.T.Chou.C.N.Yang.Phys.Rev.170.1591.1968}T. T. Chou and C. N. Yang, Phys. Rev. \textbf{170}, 1591 (1968); Phys. Lett. B\textbf{128}, 457 (1983).

\bibitem{PAN-82-134-2019}
V. A. Okorokov, Phys. At. Nucl. \textbf{82}, 134 (2019).

\bibitem{PU-185-963-2015}
V. V. Anisovich, Phys. Usp. \textbf{185}, 963 (2015).

\bibitem{TA-PRD-92-032007-2015}
R.U. Abbasi \emph{et al.} (Telescope Array Collaboration), Phys. Rev. \textbf{D92}, 032007 (2015).

\bibitem{PRL-109-062002-2012}
P. Abreu \emph{et al.} (The Pierre Auger Collaboration), Phys.
Rev. Lett. \textbf{109}, 062002 (2012).

\bibitem{FCC-CDR-V4-2018}
A. Abada \emph{et al.}, Eur. Phys. J. Special Topics \textbf{228}, 1109 (2019).

\bibitem{FCC-CDR-V3-2018}
A.~Abada \emph{et al.}, Eur. Phys. J. Special Topics \textbf{228}, 755 (2019).

    \end{thebibliography}
\end{document}